\documentclass{aa} 
\usepackage{graphics} 

\input{psfig} 

\begin{document}

\thesaurus{03	         
	  (11.19.3;      
	   11.05.2;      
           11.19.5;	 
	   09.08.1)	 
	 } 

\title{Evolution of photoionization and star formation in starbursts and \ion{H}{ii} 
galaxies} 

\author{Emmanuel Moy\inst{1}, Brigitte Rocca-Volmerange\inst{1,2}, Michel 
Fioc\inst{1,3} } 

\institute{Institut d'Astrophysique de Paris, 98\,{\it bis},
Boulevard Arago, F-75014 Paris, France 
\and
Institut d'Astrophysique Spatiale, B\^at. 121, Universit\'e Paris XI,
F-91405 Orsay, France
\and
NASA/Goddard Space Flight Center, code 685, Greenbelt, MD 20771, USA
} 

\offprints{moy@iap.fr} 

\date{} 

\authorrunning{E. Moy, B. Rocca-Volmerange, M. Fioc} 

\titlerunning{Photoionization and star formation in starbursts} 

\maketitle 

\begin{abstract}
We analyze coherently the stellar and nebular energy
distributions of starbursts and \ion{H}{ii} galaxies, using our
evolutionary synthesis model, P\'EGASE (Fioc \& Rocca-Volmerange 1997,
2000), coupled to the photoionization code CLOUDY (Ferland
1996). The originality of this study is to relate the evolution and
the metallicity of the starburst to the past star formation history of
the host galaxy. Extinction and geometrical effects on emission lines
and continua are computed in coherency with metallicity. 
We compare our model predictions to an observed sample of $\approx 750$
\ion{H}{ii} regions and starbursts.

When fitting [\ion{O}{iii}]$_{\lambda 4363}$/[\ion{O}{iii}]$_{\lambda 5007}$, [\ion{O}{i}]$_{\lambda 6300}$/H$\alpha$, 
[\ion{S}{ii}]$_{\lambda\lambda 6716, 6731}$/H$\alpha$, [\ion{N}{ii}]$_{\lambda 6584}$/H$\alpha$ and [\ion{O}{iii}]$_{\lambda 5007}$/H$\beta$, the most striking feature 
is the decreasing spread in $U$ with
increasing metallicity $Z$. High-$U$ objects systematically have a
low metallicity while low levels
of excitation happen at any $Z$. The best fits of emission line 
ratios 
are obtained with a combination of a high- and a low-ionization
components. No additional source of ionizing photons --
shocks or hidden AGN -- is needed. 
The high level of excitation observed in metal-poor \ion{H}{ii} galaxies
requires a very 
young population ($\le 3$\,Myr), while starburst nuclear galaxies (SBNGs) are 
consistent with a wider range of age ($\le 5$\,Myr). 

Colors ($B-V$, $V-R$) and equivalent widths are fitted
in coherency with emission line ratios. An underlying
population is needed, even for small-aperture observations. This evolved 
population not only reddens the
continuum and dilutes the equivalent width of the emission lines, but also 
participates
in the ionization process. Its main effect 
on line ratios is to maintain a high level of excitation
when the burst stops. Models combining underlying populations typical
of Hubble sequence galaxies and instantaneous starbursts
with ages between 0 and 8\,Myr agree satisfactorily
with all the data. 


\end{abstract}

\section{Introduction}
\begin{table*}[thbp]
\begin{center}
\begin{tabular}{cccc}
\hline References & Starburst type & Balmer absorption & Reddening \\ &
& correction & correction \\ \hline Terlevich et al. 1991 & \ion{H}{ii}G,
SBNG & no & no\\ Veilleux \& Osterbrock 1987 & SBNG & no& no\\
Veilleux et al. 1995 & IRAS SBNG & yes& no\\ French 1980 & \ion{H}{ii}G & no &
no \\ Leech et al. 1989 & IRAS SBG& yes& yes\\ Storchi-Bergmann et al.
1995 & \ion{H}{ii}G, SBNG & no& no \\ Contini et al. 1998 & IRAS SBNG, \ion{H}{ii} &
yes& no \\ \hline 
\end{tabular}
\vskip 0.5cm
\parbox{13cm}{\caption{Data sources. SBNG: starburst
nucleus galaxies; \ion{H}{ii}G: \ion{H}{ii} galaxies; \ion{H}{ii}: \ion{H}{ii} regions.} }
\end{center}
\end{table*}
The ``starburst'' phenomenon calls to mind a class of objects 
dominated by the radiation of massive stars (Gallego et al. 1995)
embedded in a dusty \ion{H}{ii} region. Modelling starbursts consistently
is the key to interpret the
properties of the actively star-forming galaxies detected at a redshift 
$z>2$ (Giavalisco 
et al. 1996; Madau et al. 1996; Steidel et al. 1996; Lowenthal et al.
1997). To this purpose, the extensive study of local samples
is a requisite step. 

Local starbursts span a wide range of types, from individual \ion{H}{ii} 
regions 
in spiral arms to blue compact \ion{H}{ii} galaxies and huge kpc-scale
nuclear starbursts (Coziol et al. 1998). Types differ from one another
in their main characteristics (line ratios, equivalent widths,
colors), and presumably have different stellar populations and physical
conditions. So, is it possible to find correlations between the basic properties: 
age, initial mass function, metallicity, relative distributions
of stars, gas and dust? Among these, which
one dominates spectral properties? In which range do parameters vary? 
Answering these questions requires
a consistent model of starbursts coupling the evolution of stars, dust and gas. 

The first studies dealing with emission lines of \ion{H}{ii} regions (Shields
1974, 1978; Stasi\'nska 1978, 1980; McCall et al. 1985)
relied on single-star photoionization models. Average physical
properties of large starburst samples were deduced from these
pioneering works. As an example, a relation between the metallicity $Z$
and the ionization parameter $U$ was derived by Dopita \& Evans
(1986) by fitting the emission line ratios from large samples of
extragalactic \ion{H}{ii} regions. 

More realistic stellar populations of star clusters were computed by McGaugh (1991) with
 the Salpeter (1955) initial mass function (IMF), to calibrate metallicity dependant line ratios. 
Further improvements were the implementation of theoretical 
tracks of massive stars, allowing to follow the evolution in time of a starburst  (e.g.
Garc\'{\i}a-Vargas \& D\'{\i}az 1994).  Thanks to the computation of 
evolutionary tracks for various metal abundances, the 
influence of the metallicity could also be studied (Cid-Fernandes et al. 1992; Cervi\~no \& 
Mas-Hesse 1994; Olofsson 1997; Garc\'{\i}a-Vargas et al. 1995a,b; Stasi\'nska \& 
Leitherer 1996). Evans (1991) emphasized the importance of using up-to-date stellar atmosphere models to calibrate nebular diagnostics. The most 
recent models take into account modern, updated stellar physics
(Leitherer \&
Heckman 1995; Leitherer et al. 1999), in particular the impact  on stellar spectra of line blanketing and of departures 
from local thermodynamic equilibrium   (Stasi\'nska \& Schaerer 1997).

Till now, only a few large datasets
covering a wide range of physical conditions
have been analyzed with state-of-the-art models, coupling evolving
stellar populations and 
photoionization. Stasi\'nska \& Leitherer (1996), for example,
restricted their analysis to metal-poor ($Z<Z_{\odot}/4$)
objects. As a matter of fact, statistical properties of starbursts 
are not definitely established. A
metal-dependent IMF was early proposed as an explanation for the
[\ion{O}{iii}]$_{\lambda 5007}$/H$\beta$ decrease at high $Z$ (Terlevich 1985;
Shields \& Tinsley 1976). Although a standard IMF seems in agreement
with the bulk of observations (Leitherer 1998), the IMF slope (see e.g.
Eisenhauer et al. 1998; Greggio et al. 1998) and mass cut-offs (Goldader
et al. 1997) are still under debate. 

Our aim is hereafter to study the evolution of the spectral properties of
aging starbursts with the help of models taking into account the effects
of metallicity, geometry, dust and the excitation level of the gas. 
To avoid normalization problems, we 
selected relative properties independent of 
distance (colors, equivalent widths and line ratios). We focused on emission 
line ratios at 
close wavelengths -- thus reducing reddening effects -- 
to study the ionizing 
spectrum and the excitation of the gas. Equivalent widths and colors are used 
preferentially to study age effects, as well as the impact of the spatial 
distribution of stars, gas and dust. 

The observational sample is presented in Sect.~2. The 
coupling of the codes P\'EGASE, to model the evolution of star formation 
and stellar emission, and CLOUDY, for a consequent photoionization of the gas by 
massive stars in a given geometry,
is described in Sect.~3. The relation between the metallicity $Z$ and the 
ionization parameter $U$ is analyzed in Sect.~4, while equivalent widths and 
colors are considered in Sect.~5.
The contribution of an underlying population is analyzed in Sect.~6. 
Discussion and conclusion are respectively in Sect.~7 and~8. 
\section{A selected sample of starbursts and \ion{H}{ii} galaxies}
\begin{table*}[thbp]
\begin{center}
\begin{tabular}{|c|c|c|}
\hline
Line & P\'EGASE+CLOUDY & Garc\'{\i}a-Vargas et al. (1995) \\
\hline
$\log(\mathrm{H}\beta)$ [erg\,s$^{-1}$] & 38.83 & 38.84 \\
\hline
[\ion{O}{ii}]$_{\lambda 3727}$/H$\beta$ & 3.10 & 3.07 \\
\hline
[\ion{O}{iii}]$_{\lambda 5007}$/H$\beta$ & 0.31 & 0.25 \\
\hline
[\ion{O}{i}]$_{\lambda 6300}$/H$\beta$ & 0.04 & 0.04 \\
\hline
[\ion{N}{ii}]$_{\lambda 6584}$/H$\beta$ & 1.34 & 1.34 \\
\hline
[\ion{S}{ii}]$_{\lambda 6716}$/H$\beta$ & 0.57 & 0.58 \\
\hline 
[\ion{S}{ii}]$_{\lambda 6731}$/H$\beta$ & 0.39 & 0.40 \\
\hline
[\ion{S}{iii}]$_{\lambda 9069}$/H$\beta$ & 0.36 & 0.34 \\
\hline 
\end{tabular}
\vskip 0.25cm
\begin{tabular}{|c|c|c|}
\hline
Line & P\'EGASE + CLOUDY & Stasi\'nska \& Leitherer (1996) \\
\hline
$\log(\mathrm{H}\beta)$ [erg\,s$^{-1}$] & 40.56 & 40.56 \\
\hline
[\ion{O}{ii}]$_{\lambda 3727}$/H$\beta$ & 2.57$\times 10^{-1}$ & 2.75$\times 10^{-1}$ \\
\hline
[\ion{O}{iii}]$_{\lambda 5007}$/H$\beta$ & 8.18$\times 10^{-1}$ & 7.37$\times 10^{-1}$ \\
\hline
[\ion{O}{i}]$_{\lambda 6300}$/H$\beta$ & 2.60$\times 10^{-3}$ & 3.13$\times 10^{-3}$ \\
\hline
\end{tabular}
\vskip 0.5cm
\parbox{13cm}{\caption{Comparison with  predictions of coupled models
available in the literature. Top: Garc\'{\i}a-Vargas et al. (1995);
$\mathrm{age}=1\,\mathrm{Myr}$, $Z=Z_{\odot}$, $\log(U)=-3.11$. 
Bottom: Stasi\'nska \& Leitherer (1996); $\mathrm{age}=1\,\mathrm{Myr}$, 
$Z=Z_{\odot}$, $\log(U)=-2.60$.}} 
\end{center}
\end{table*}
Our data set (Table~1) is a selection of 754 objects from observational 
samples of galaxies having properties characteristic of starbursts: 
intense emission lines (French 1980; Veilleux \&
Osterbrock 1987; Terlevich et al. 1991), strong far-infrared emission
(Leech et al. 1989; Veilleux et al. 1995), or  excess of blue or
ultraviolet emission (Storchi-Bergmann et al. 1995); in addition, 
Contini et al. (1998) require galaxies to have a bar. Objects 
dominated by an active galactic nucleus (AGN) are excluded from our 
sample, though some of them may be faintly contaminated by nuclear activity. 
Effects of aperture and differences in data reduction processes are considered 
in Sect.~2.2.
\subsection{Spectral classification of the original sample}
Objects listed in the source papers are classified in two
families. The first one includes metal-poor \ion{H}{ii} galaxies
(\ion{H}{ii}Gs) and extra-nuclear \ion{H}{ii} regions, both likely devoid of active
nucleus. The second one gathers galactic central regions where the ionized-gas 
emission
could be partly due to shocks (Kim et al. 1998)
or AGNs (see Heckman 1991 and references therein). To select only 
starburst nucleus galaxies (SBNGs) in this last class, we
keep the classification of the authors, which is based 
on the pioneering works of Baldwin et al.
(1981, hereafter BPT) and Veilleux \& Osterbrock (1987, hereafter VO).

The Terlevich et al. (1991) sample is mainly composed of \ion{H}{ii}Gs
with some possible SBNGs. The
authors used two of the BPT criteria based on the [\ion{O}{iii}]$_{\lambda 5007}$/H$\beta$ and [\ion{O}{ii}]$_{\lambda 3727}$/[\ion{O}{iii}]$_{\lambda 5007}$ ratios. We
also analyzed their sample with the VO method and obtained the same
classification. 

Contini et al. (1998) adopted the VO criteria. Their sample includes
SBNGs and giant \ion{H}{ii} regions located in IRAS barred spiral galaxies,
selected from Mazzarella \& Balzano (1986) and listed in the
Lyon-Meudon Extragalactic Database (LEDA). 

Two samples contain only SBNGs located inside IRAS galaxies. Veilleux
et al. (1995) followed both BPT and VO criteria, while 
Leech et al. (1989) used only some of the criteria of BPT. Actually, an
important part of their sample lies far beyond the limit reported by VO 
between
\ion{H}{ii}-like regions and AGNs in the [\ion{O}{iii}]/H$\beta$ vs. [\ion{O}{i}]/H$\alpha$ and [\ion{O}{iii}]/H$\beta$ vs. [\ion{N}{ii}]/H$\alpha$ diagrams. These objects are clearly misclassified and have been 
excluded from our selection. Storchi-Bergmann et al.
(1995) reported observations of emission line galaxies of various
types. Following Coziol et al. (1998), all compact galaxies are hereafter 
classified as \ion{H}{ii} galaxies, whether dwarf or not.
Finally, we include in our
selection the observations of five starbursts from the sample of
Balzano (1983) reported by VO, and the data of French (1980) on 14
\ion{H}{ii}Gs. 
\subsection{Homogeneity of reddening, aperture and absorption corrections}
Most line ratios considered here are computed from close lines; so, reddening 
 effects should be small and will be neglected in the following.

In most samples, the 
contamination, through large apertures, of the starburst light by the host 
galaxy population is likely to increase the continuum emission and to 
dilute the equivalent width of emission
lines. To avoid this problem, we will restrict the analysis of colors and 
equivalent widths to the sample of Contini et al.
(1998). These data were acquired with a long-slit spectrograph, but the spectra presented by the authors correspond to H$\alpha$ emitting regions exclusively. Hence, the 
contamination of the starburst continuum by the environment of the host galaxy 
should be very 
weak. 

The host galaxy can also modify the apparent line ratios involving Balmer lines 
through the presence of stellar absorption lines. 
Emission line fluxes are corrected for this effect as follows: 
\begin{equation}
F^\mathrm{corr}_\mathrm{line}=F^\mathrm{obs}_\mathrm{line}\left(1 + 
\frac{W^\mathrm{abs}_\mathrm{line}}{W^\mathrm{obs}_\mathrm{line}}\right), 
\end{equation}
where $F^\mathrm{corr}_\mathrm{line}$ and $F^\mathrm{obs}_\mathrm{line}$ 
are respectively the
absorption-corrected and the apparent emission line fluxes, and
$W^\mathrm{abs}_\mathrm{line}$ 
and $W^\mathrm{obs}_\mathrm{line}$ are the absorption and measured emission 
equivalent widths. 
Not all the samples were corrected (Table~1). The correction should be 
negligible for the H$\alpha$ line because $W^\mathrm{abs}_{\mathrm{H}\alpha}$
$\ll$ $W^\mathrm{obs}_{\mathrm{H}\alpha}$. It is stronger for H$\beta$ 
(up to 100\,\% in Leech et al. 1989) and
the ratio [\ion{O}{iii}]/H$\beta$ may be affected. This ratio is systematically higher in the samples of French (1980) and Terlevich et al. (1991), mainly composed of HIIGs, than in the other ones, which  include mainly SBNGs. Is this trend real, or is it due to the fact that these two data sets are not corrected for underlying stellar lines? Note that Storchi-Bergmann et al. 
(1995) already noticed that [\ion{O}{iii}]/H$\beta$ is higher in \ion{H}{ii}
galaxies than in SBNGs. Moreover, the correction for absorption lines should be very small in \ion{H}{ii}Gs, since these objects have simultaneously large $W^\mathrm{obs}_{\mathrm{H}\beta}$ (67\,\AA\ in Terlevich et al. 1991) and low  $W^\mathrm{abs}_{\mathrm{H}\beta}$. Hence, we conclude that the trend we observe is real and not due to the discrepancies between the reduction procedures.

\section{The coupling of spectral evolution and photoionization} 
\begin{center}
\begin{table*}[htbp]
\begin{center}
\begin{tabular}{|c|c|l|}
\hline
& & \\
{Element X} & {(X/H)$_{Z_{\odot}}$} & \hskip 1.5cm {(X/H)$_{Z}$} \\
& & \\
\hline
He & 0.107 & $0.08096+0.01833 \times Z/(0.7 Z_{\odot})$ \\
\hline
B & 2.63$\times 10^{-11}$ & (B/H)$_{Z_{\odot}} \times Z/Z_{\odot}$ \\
\hline
C & 3.63$\times 10^{-4}$ & (C/H)$_{Z_{\odot}} \times Z/Z_{\odot}$ \\
\hline
O & 8.51$\times 10^{-4}$ & (O/H)$_{Z_{\odot}} \times Z/Z_{\odot}$ \\
\hline
F & 3.63$\times 10^{-8}$ & (F/H)$_{Z_{\odot}} \times Z/Z_{\odot}$ \\
\hline
Na & 2.14$\times 10^{-6}$ & (Na/H)$_{Z_{\odot}} \times Z/Z_{\odot}$ \\
\hline
P & 2.82$\times 10^{-7}$ & (P/H)$_{Z_{\odot}} \times Z/Z_{\odot}$ \\
\hline
Cl & 3.16$\times 10^{-7}$ & (Cl/H)$_{Z_{\odot}} \times Z/Z_{\odot}$ \\
\hline
Ar & 3.63$\times 10^{-6}$ & (Ar/H)$_{Z_{\odot}} \times Z/Z_{\odot}$ \\
\hline
Fe & 4.86$\times 10^{-5}$ & (Fe/H)$_{Z_{\odot}} \times Z/Z_{\odot}$ \\
\hline
\end{tabular}
\vskip 0.5cm 

$\log(\mathrm{N}/\mathrm{O})$\\ 

\begin{tabular}{|c|}
\hline
$12.43-3.77 (\log(\mathrm{O}/\mathrm{H}) + 12) + 0.26 (\log(\mathrm{O}/\mathrm{H})+12)^2$ \\
\hline 
\end{tabular}
\vskip 0.5cm
\parbox[b]{11cm}{\caption{Adopted solar abundances and prescriptions for 
abundance variations with $Z$}} 
\end{center}
\end{table*}
\end{center}
The evolutionary synthesis code we use, P\'EGASE, 
takes into account metallicity and dust
effects. P\'EGASE computes the stellar spectral energy distributions (SEDs) and 
the metallicities of
starbursts and galaxies of the Hubble sequence at any stage
of evolution, within the metallicity range $Z=10^{-4}$
to $10^{-1}$. 
Typical parameters of P\'EGASE are the star
formation rate (SFR) and the initial mass function. The new 
version\footnote{This code is available at 
\emph{http://www.iap.fr/users/fioc/PEGASE.html} or by anonymous ftp
at \emph{ftp.iap.fr} in \emph{/pub/from$\_$users/pegase}.}
used 
hereafter, P\'EGASE.2, is based on the evolutionary tracks of Girardi et al. 
(1996), Fagotto et al. (1994a,b,c) and Bressan et al. (1993). The AGB to 
post-AGB phases are computed following the prescriptions of 
Groenewegen \& de Jong (1993) models. 
The synthetic stellar spectral library is 
from Kurucz (1992), modified by Lejeune et al. (1997) to fit observed colors. 
For details, see Fioc \& Rocca-Volmerange (1999; 2000, in preparation)

The photoionization code CLOUDY (version 90.04, Ferland 1996) predicts the 
spectra of
low- to high-density astrophysical plasmas in the
most extreme astrophysical sites, such as starbursts and quasar environments. 
It takes into account recent changes in 
atomic databases and new numerical methods. In addition to its physical
performances, the code was chosen for its easy Web access, its
documentation and its friendly on-line help. CLOUDY proposes options to 
explore the influence of geometrical factors: type of geometry --
spherical or plane parallel --, covering factor 
$\Omega/(4\pi)$, filling factor $f$, and the distance $R$ from the source to 
the plasma.

The coupling of P\'EGASE.2 with CLOUDY allows to compute coherently the
stellar and nebular emission, and to fit  simultaneously
the continuum and lines from
stellar populations embedded in a gas cloud. The metallicity $Z(t)$, 
and the stellar SED -- in particular the energy distribution of the ionizing 
photons -- provided by P\'EGASE are used as inputs by CLOUDY. 
Hereafter, we call $Q_\mathrm{H^{0}}$ the emission rate of
ionizing photons. This quantity is used to normalize the  SEDs computed by P\'EGASE.

The results of the current version of the code have been compared to those 
obtained by similar ``coupled models''. They are presented in Table~2 
for two sets of predictions: the model 16 from Garc\'{\i}a-Vargas et 
al. (1995b), and the $Z_{\odot}$-model computed by Stasi\'nska \& Leitherer 
(1996) for a $10^{6}\,M_{\odot}$-cluster. In both cases, we used 
inputs matching the reference model (IMF, age, geometrical parameters 
and chemical abundances). For the first comparison test, the agreement is very 
good (Table~2, top). Our results also agree relatively well with those of 
Stasi\'nska \& Leitherer (1996), given that the evolutionary tracks, the 
spectral libraries and the photoionization codes are different in the two 
models.

\

\subsection{Star formation parameters}
An instantaneous star formation episode
(``pure starburst'') is assumed for starbursts -- a scenario 
well suited to study emission line properties.
In Sect.~6, we will also consider an underlying stellar population
formed either during a previous burst or continuously.
We tentatively tested several IMFs
of the form d$N$/d$m\propto m^{-\alpha}$, $m\in[M_\mathrm{low}, 
M_\mathrm{up}]$: $\alpha=-2.35$, $m\in[0.1, 120]\,M_{\odot}$ (Salpeter 1955);
$\alpha=-3.35$, $m\in[0.1, 120]\,M_{\odot}$;
and $\alpha=-2.35$, $m\in[3, 30]\,M_{\odot}$.
The last one is truncated, as suggested by various infrared studies 
(Rieke et al. 1993; Lan\c{c}on \&
Rocca-Volmerange 1996). The initial metallicity
is left as a free parameter. 
\subsection{Physical properties of the gas}
\begin{center}
\begin{figure*}[htbp]
\begin{center}
\centerline{\hbox{
\psfig{figure=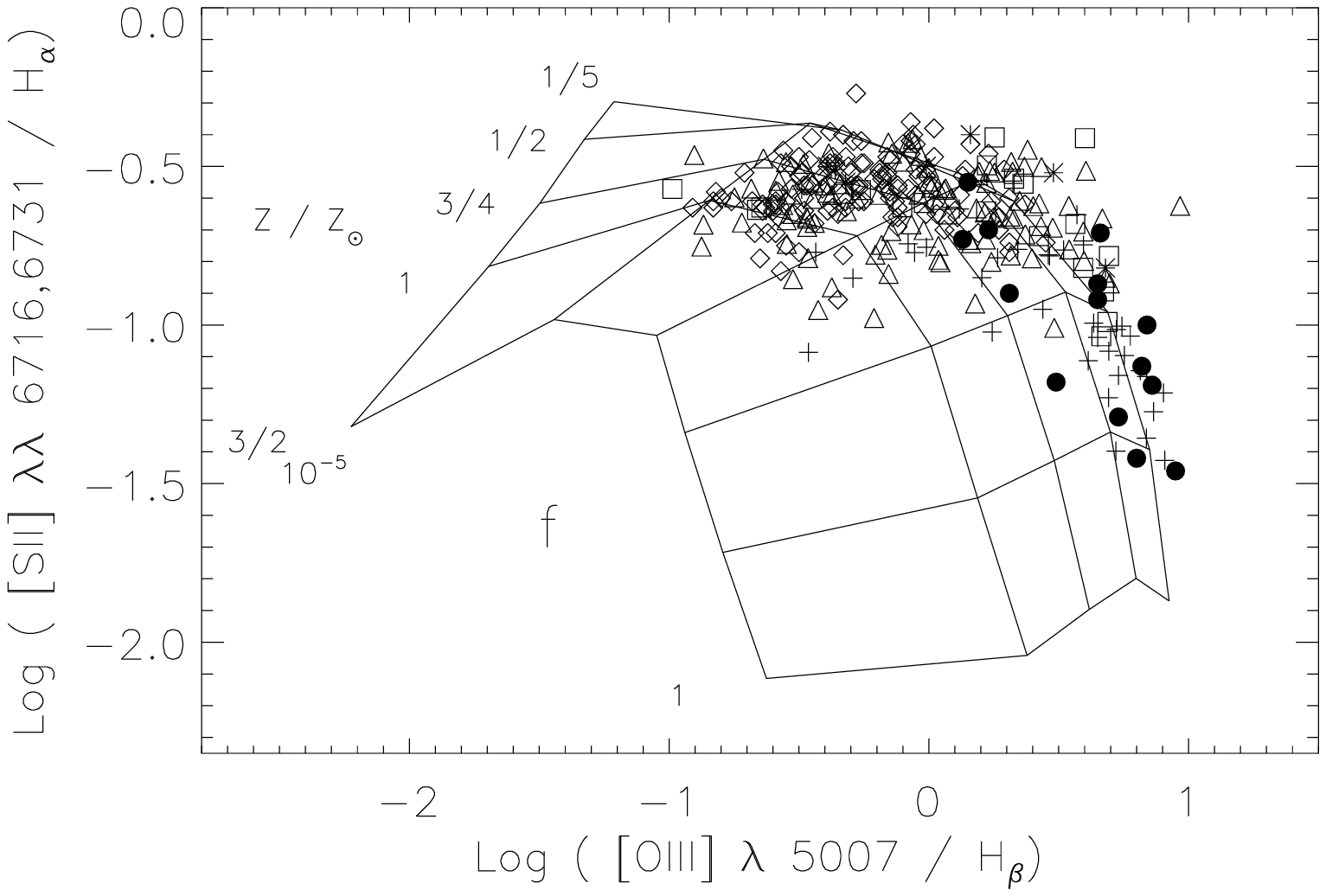,bbllx=20pt,bblly=10pt,bburx=480pt,bbury=312pt,width=7.5cm}
\psfig{figure=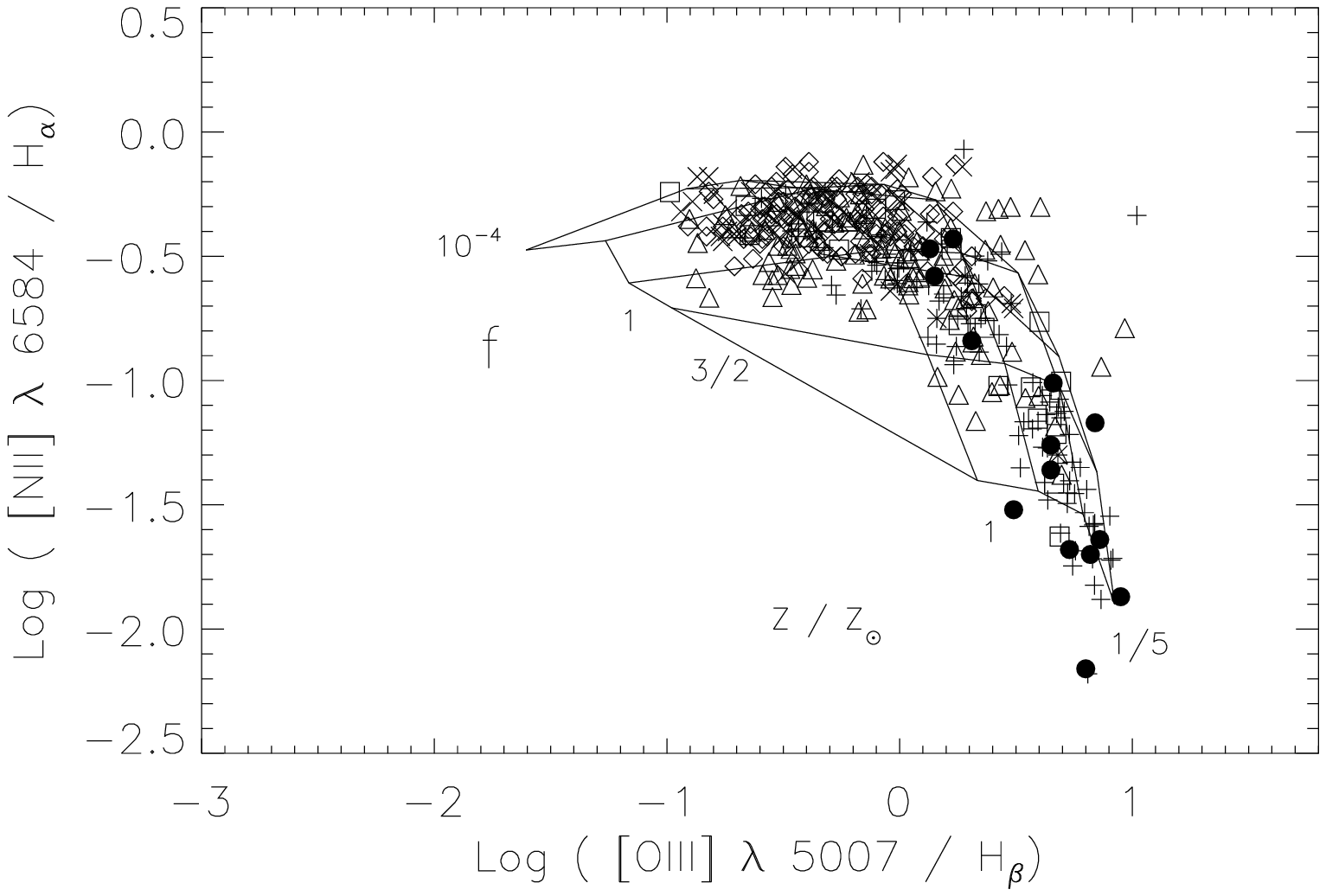,bbllx=20pt,bblly=10pt,bburx=480pt,bbury=312pt,width=7.5cm}}}
\centerline{\hbox{
\psfig{figure=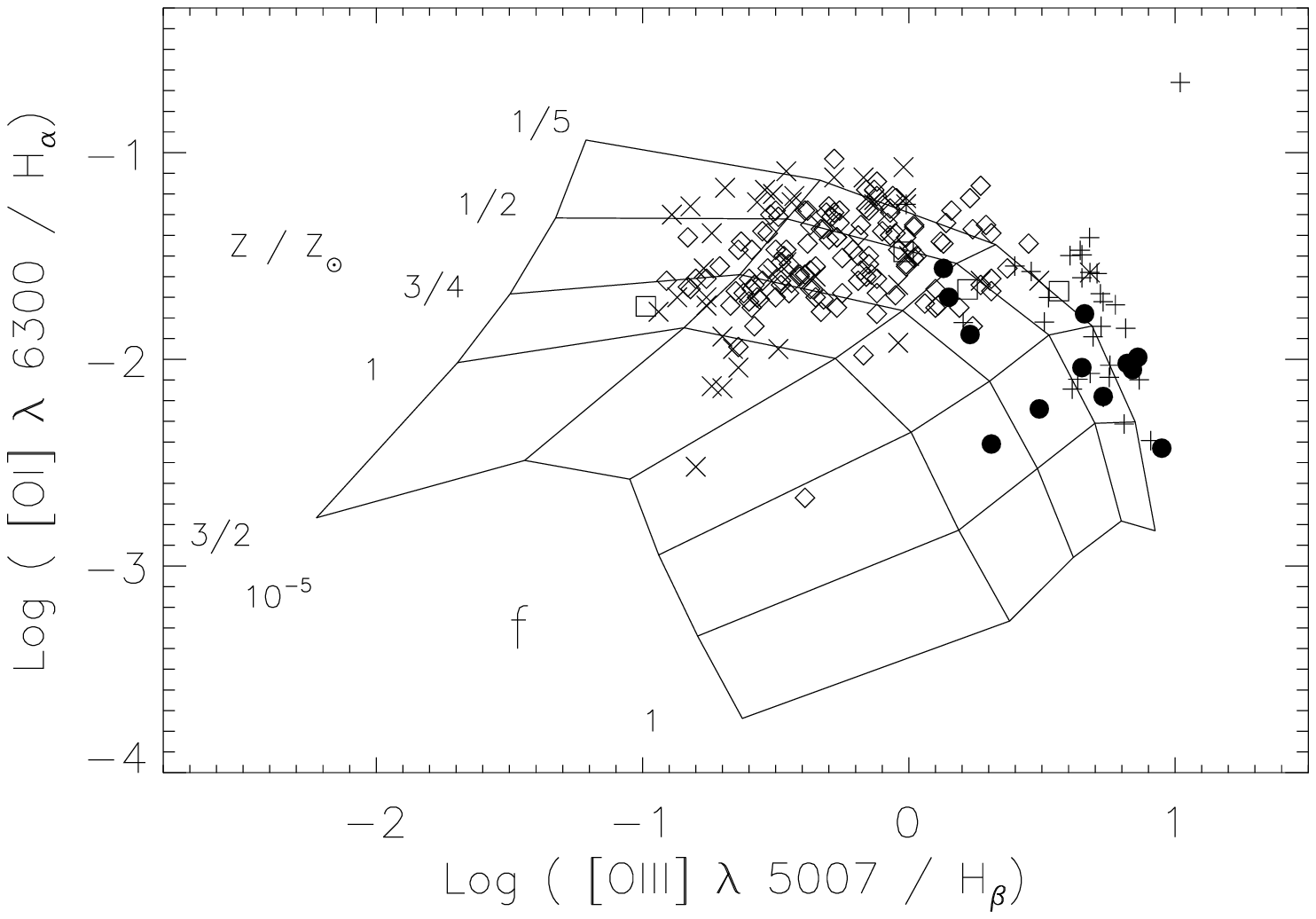,bbllx=20pt,bblly=10pt,bburx=480pt,bbury=312pt,width=7.5cm}
\psfig{figure=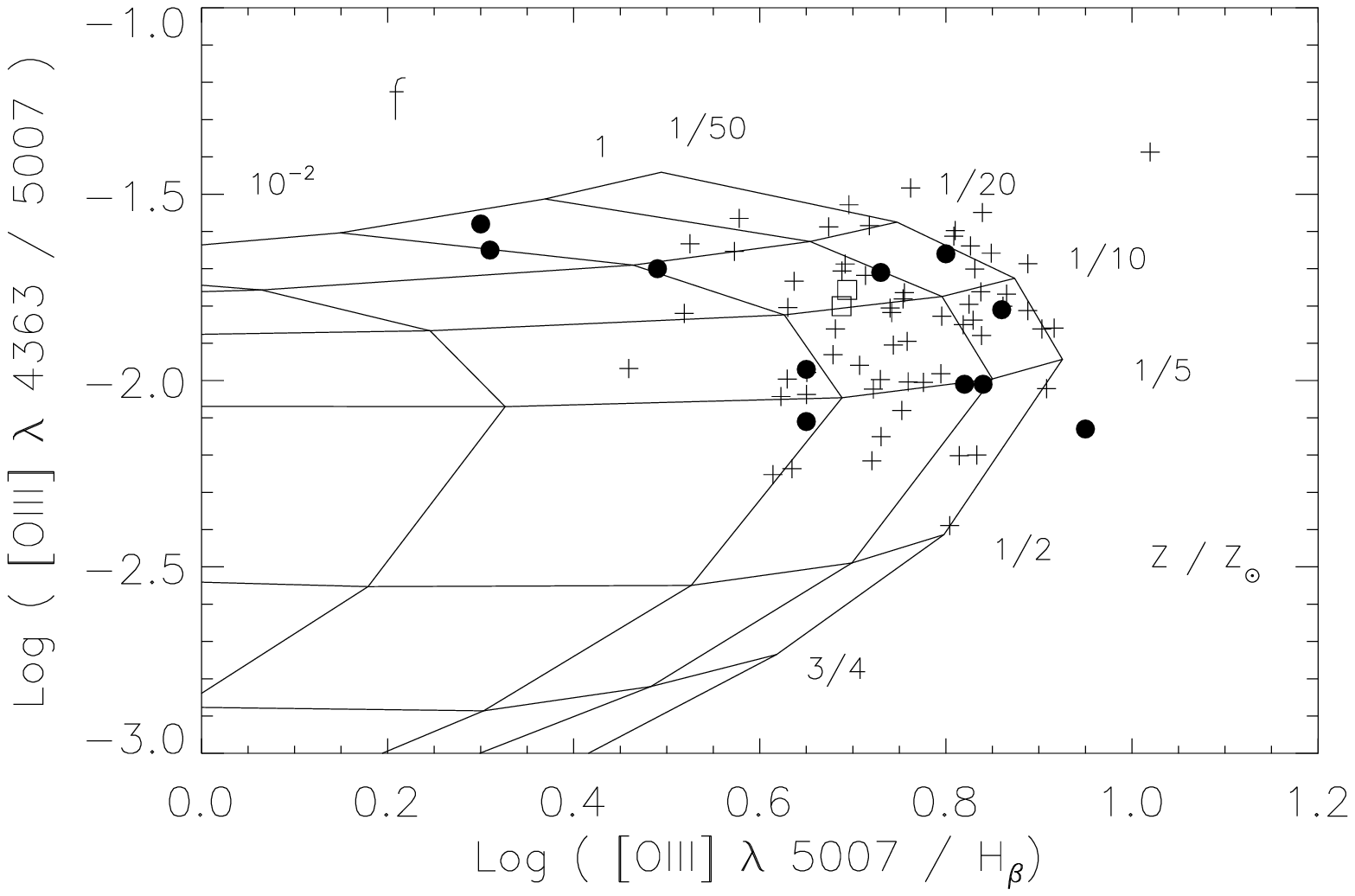,bbllx=20pt,bblly=10pt,bburx=480pt,bbury=312pt,width=7.5cm}
}}
\parbox{15cm} {\caption{Models of pure instantaneous starbursts for
various filling factors ($f$) and metallicities ($Z$). The data are from
Terlevich et al. (``plus'' signs), Veilleux \& Osterbrock (asterisks), French 
(filled circles), Veilleux et al. (diamonds), Contini et al. (triangles),
Storchi-Bergmann et al. (squares) and Leech et al. (crosses).
\label{GridsfZ1} }} 
\end{center}
\end{figure*}
\end{center} 
The emission line spectrum of an ionized nebula depends on the combination 
of the ionizing spectrum, the chemical composition of the gas and the so-called 
ionization parameter $U$. 

The elemental abundances of the gas have been made consistent with the metallicity 
of the ionizing stars. The solar abundances used here are given 
in Table~3. All the
abundances were scaled linearly according to the metallicity, except
for nitrogen and helium. For the likely secondary 
element N, we adopted the law proposed 
by Coziol et al. (1999), multiplied by a factor 1.5 to account for the excess 
of nitrogen observed in starbursts. For He, we followed a prescription proposed
by Dopita \& Kewley (private communication; see Table 3 for details). 

\

The ionization parameter $U$ is defined as:
\begin{equation}
U=\frac{Q_\mathrm{H^{0}}}{4 {\mathrm \pi} R_\mathrm{s}^{2}n_\mathrm{H} c}, 
\end{equation}
where $n_\mathrm{H}$ is the hydrogen density, $c$ the speed of light, and 
$R_\mathrm{s}$ the Str\"omgren radius (i.e. the radius of the ionization front). 
Taking into account the inner radius $R$ (e.g. Evans \& Dopita, 1985), 
$R_\mathrm{s}$ is roughly equal to:
\begin{equation}
R_\mathrm{s}=\left(\frac{3 Q_\mathrm{H^{0}}}{4 \pi \alpha_\mathrm{B} n_\mathrm{H}^{2}
f} + R^{3} \right)^{1/3},
\end{equation}
where $\alpha_{\mathrm{B}}$ is the case B temperature-dependent recombination 
coefficient and $f$ is the filling factor. This formula highlights
the impact of each parameter on $U$, but is only approximate, since it assumes a 
constant temperature through all the nebula. The effective Str\"omgren radius 
is obtained only at the end of a CLOUDY calculation.

The spherical geometry (Stasi\'nska \& Leitherer
1996; Gonz\'alez-Delgado et al. 1999) and plane-parallel geometry
(Garc\'{\i}a-Vargas \& D\'{\i}az 1994; Garc\'{\i}a-Vargas et al. 1997) 
are
extreme situations for which the right-hand side of eq.~(3) is
respectively dominated by the first and the second term. The adopted geometry 
was spherical. The hydrogen density $n_\mathrm{H}$
was 100\,cm$^{-3}$ in most models, but we also computed a few cases for 
$n_\mathrm{H}=10$ and $n_\mathrm{H}=1000$\,cm$^{-3}$, to assess the specific 
impact of density on emission line ratios. We set the number of ionizing 
photons to
$Q_\mathrm{H^{0}}=10^{52}$\,photons\,s$^{-1}$, corresponding to a maximum
luminosity of $\sim$ $10^{40}$\,erg\,s$^{-1}$ in the H$\alpha$ emission line 
-- a value characteristic of intense star formation sites (Shields
1990). For emission lines, the exact value of $Q_\mathrm{H^{0}}$ is not 
important as long as the density does not exceed the critical limit of 
collisional de-excitation, below which models with the same $U$, ionizing 
spectrum and metallicity predict the same line ratios. 
To obtain reasonable values for $U$, we varied the volume filling factor
between $10^{-5}$ and 1 and set $R$ to 4\,pc.

\begin{center}
\begin{figure*}[thbp]
\begin{center}
\centerline{\hbox{
\psfig{figure=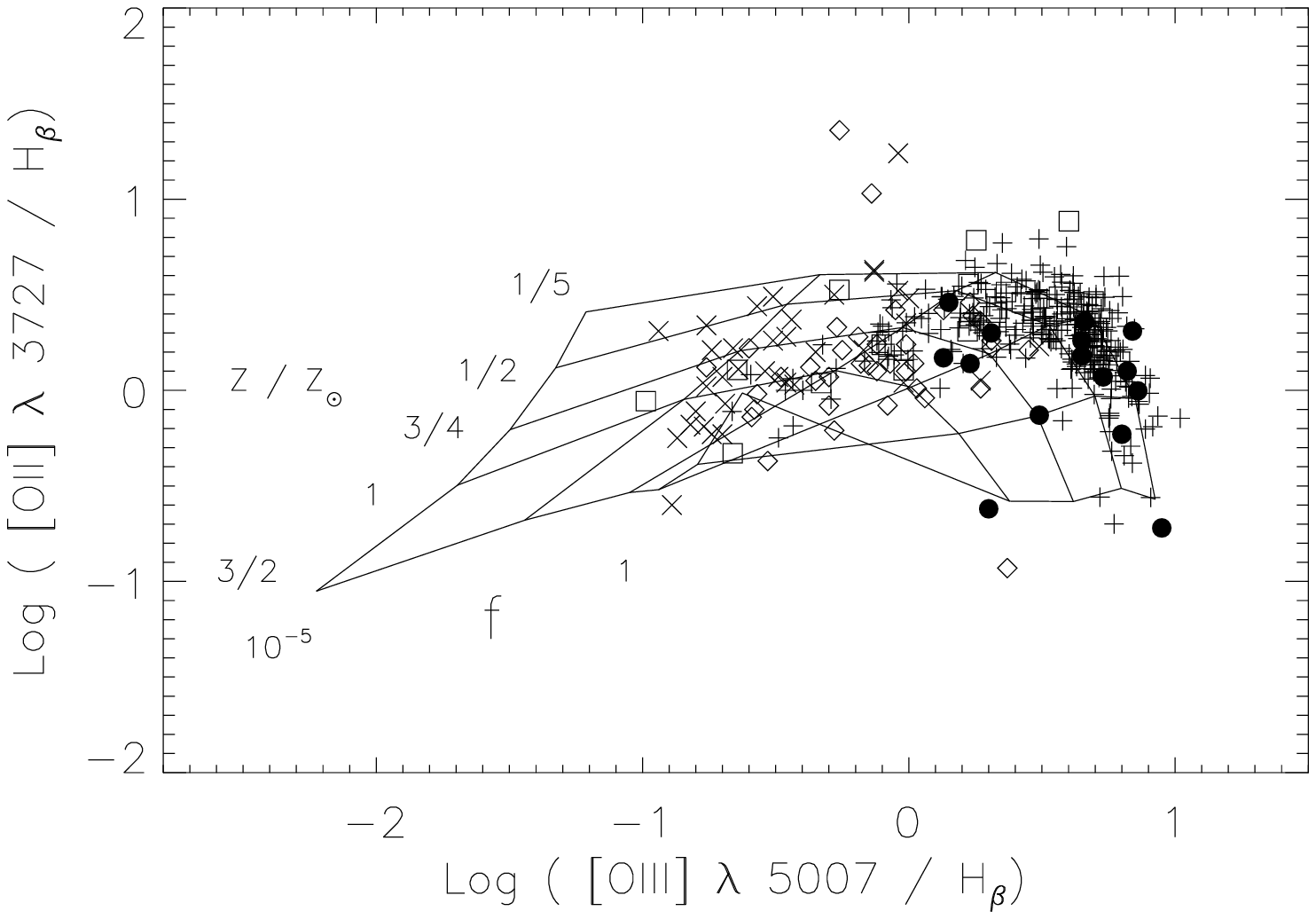,bbllx=20pt,bblly=10pt,bburx=480pt,bbury=312pt,width=7.5cm}
\psfig{figure=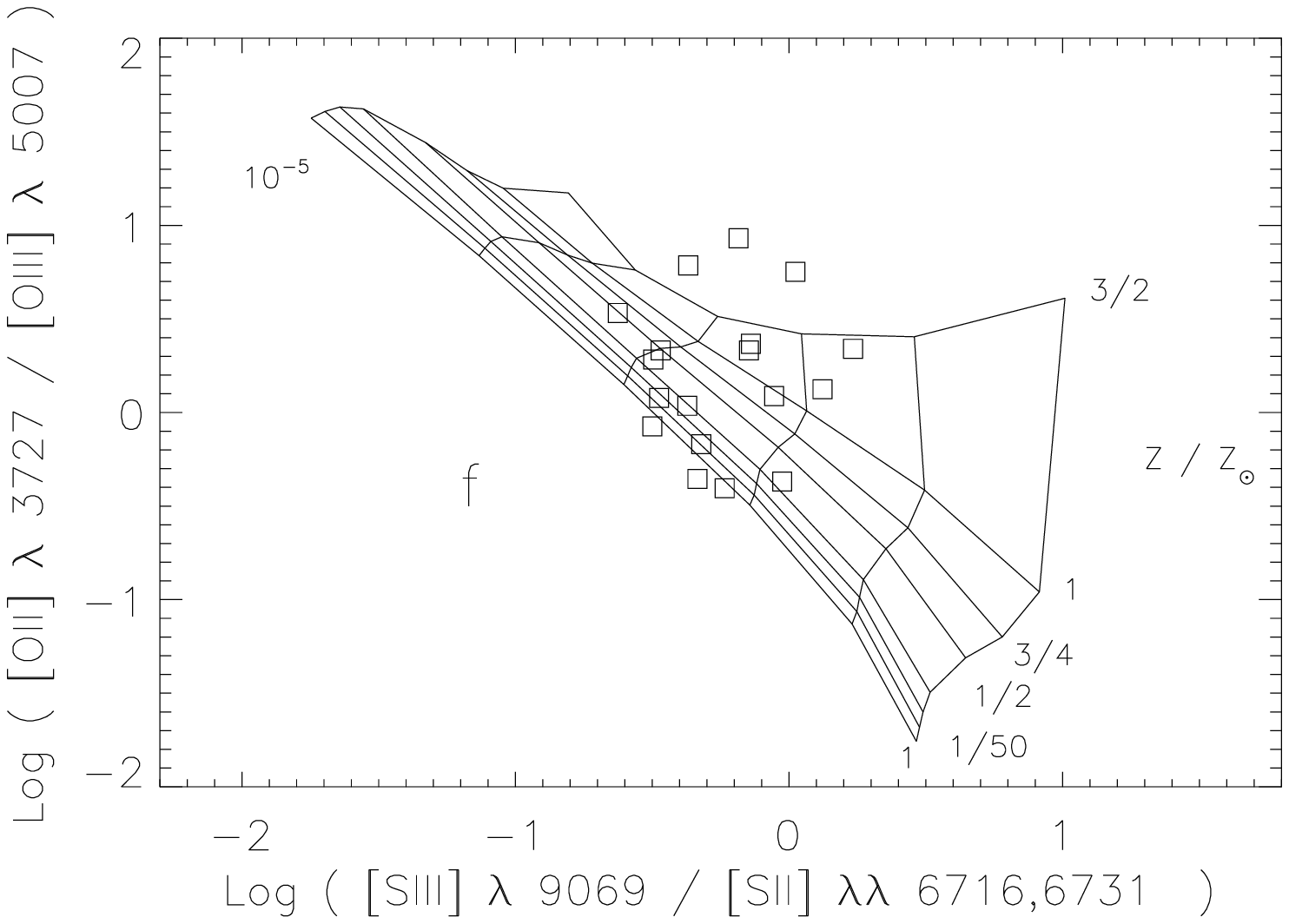,bbllx=20pt,bblly=10pt,bburx=480pt,bbury=312pt,width=7.5cm}}}
\parbox{8.3cm} {\caption{Same as Fig.~\ref{GridsfZ1} for line ratios sensitive 
to reddening. \label{GridsfZ2} }} 
\end{center}
\end{figure*}
\end{center} 

We assume that 
the gas is distributed in a spherical shell covering a solid angle $\Omega$ 
around the star cluster. The covering factor $\Omega/(4\pi)$ was allowed to 
vary in the interval [0.1, 1]. The continuum 
luminosity from the starburst $F_{\lambda}$ (erg\,s$^{-1}$ \AA$^{-1}$) at any
wavelength $\lambda$ is: 
\begin{equation}
F_{\lambda}=F_{\lambda}^\mathrm{trans} +
F_{\lambda}^\mathrm{neb} + F_{\lambda}^\mathrm{stell}. 
\end{equation}
The transmitted stellar continuum $F_{\lambda}^\mathrm{trans}$ and the
nebular continuum $F_{\lambda}^\mathrm{neb}$ both scale with 
$\Omega/(4\pi)$. $F_{\lambda}^\mathrm{stell}$, scaling as $(1-\Omega/(4\pi))$,
refers to the part of the
stellar spectrum not crossing the gas (and thus not contributing to the
ionization process). 
\subsection{Extinction of the spectrum}
Assuming a dust screen at the outer edge of the \ion{H}{ii} region, 
the continuum flux is reddened as follows: 
\begin{equation}
F_{\lambda}=(F_{\lambda}^\mathrm{trans} +
F_{\lambda}^\mathrm{neb}) e^{- \tau_{\lambda}^\mathrm{neb}} + 
F_{\lambda}^\mathrm{stell} e^{-\tau_{\lambda}^\mathrm{stell}}. 
\end{equation} 
Emission line fluxes are also multiplied by
$e^{- \tau_{\lambda}^\mathrm{neb}}$. 
The optical depth $\tau_{V}^\mathrm{neb}$ was a free
parameter, as well as the 
$\tau_{V}^\mathrm{stell}/ \tau_{V}^\mathrm{neb}$ ratio ($\leq 1$). This option 
takes
into account a possible dust concentration outside the
nebula rather than on the line-of-sight not covered by the gas. 
The variation of the optical depth with wavelength is derived from
the metallicity-dependent extinction laws of the Magellanic Clouds and 
our Galaxy.
\section{Results on emission line ratios for instantaneous starburst models}

We have explored the impact of the IMF, the age of the starburst, the metallicity and the ionization parameter on the following optical line ratios: [\ion{O}{iii}]$_{\lambda 4363}$/[\ion{O}{iii}]$_{\lambda 5007}$, 
[\ion{O}{i}]$_{\lambda 6300}$/H$\alpha$, [\ion{S}{ii}]$_{\lambda\lambda 6716, 6731}$/H$\alpha$, 
[\ion{N}{ii}]$_{\lambda 6584}$/H$\alpha$ and [\ion{O}{iii}]$_{\lambda 5007}$/H$\beta$.  The value of $f$ was varied between 
$10^{-5}$ and 1. 
The corresponding $\log(U)$ (as
defined in Sect.~3.2) belong to $[-5.5, -1.5]$. Note that similar results could
also be derived by varying the inner radius $R$ (in a plane-parallel 
geometry), the density $n_\mathrm{H}$ or a combination of both.

\

\subsection{Metallicity effects}
Emission line ratios are sensitive to the metallicity through two competing 
effects.
On the one hand, the opacity increases with $Z$, lowering 
the stellar effective temperatures and thus
the hardness of the ionizing spectrum. A trend towards lower excitation levels 
is therefore expected in high-$Z$
starbursts, even if there is no dependence of the
IMF or the geometry on the metallicity. This effect is reinforced at high $Z$ by the
efficient cooling by metals and thus the low electronic temperature 
$T_\mathrm{e}$.

On the other hand, the energy in metal emission lines tends
naturally to increase
with higher elemental abundances. For oxygen lines, this
process dominates at $Z\leq 1/5 Z_{\odot}$, while the opposite happens at higher metallicity (e.g. McGaugh 1991). As a consequence,
two models with metallicities respectively below and above this limit
can predict comparable emission lines. For this reason, we present hereafter 
our predictions for $Z$ in the range $[1/50, 3/2] Z_{\odot}$ if no degeneracy 
happens in the considered diagram, but in $[1/5, 3/2] Z_{\odot}$ otherwise. 
\subsection{Effects of the geometrical filling factor $f$}

For a given density $n_\mathrm{H}$, varying the filling factor $f$ changes 
the optical depth per unit length and the ionization structure: the smaller 
$f$, the smaller the mean ionization level, and the
lower the luminosities of high excitation lines such as 
[\ion{O}{iii}]$_{\lambda 5007}$. Low-excitation lines behave inversely. 

\begin{figure*}[htbp]
\begin{center}
\centerline{\hbox{
\psfig{figure=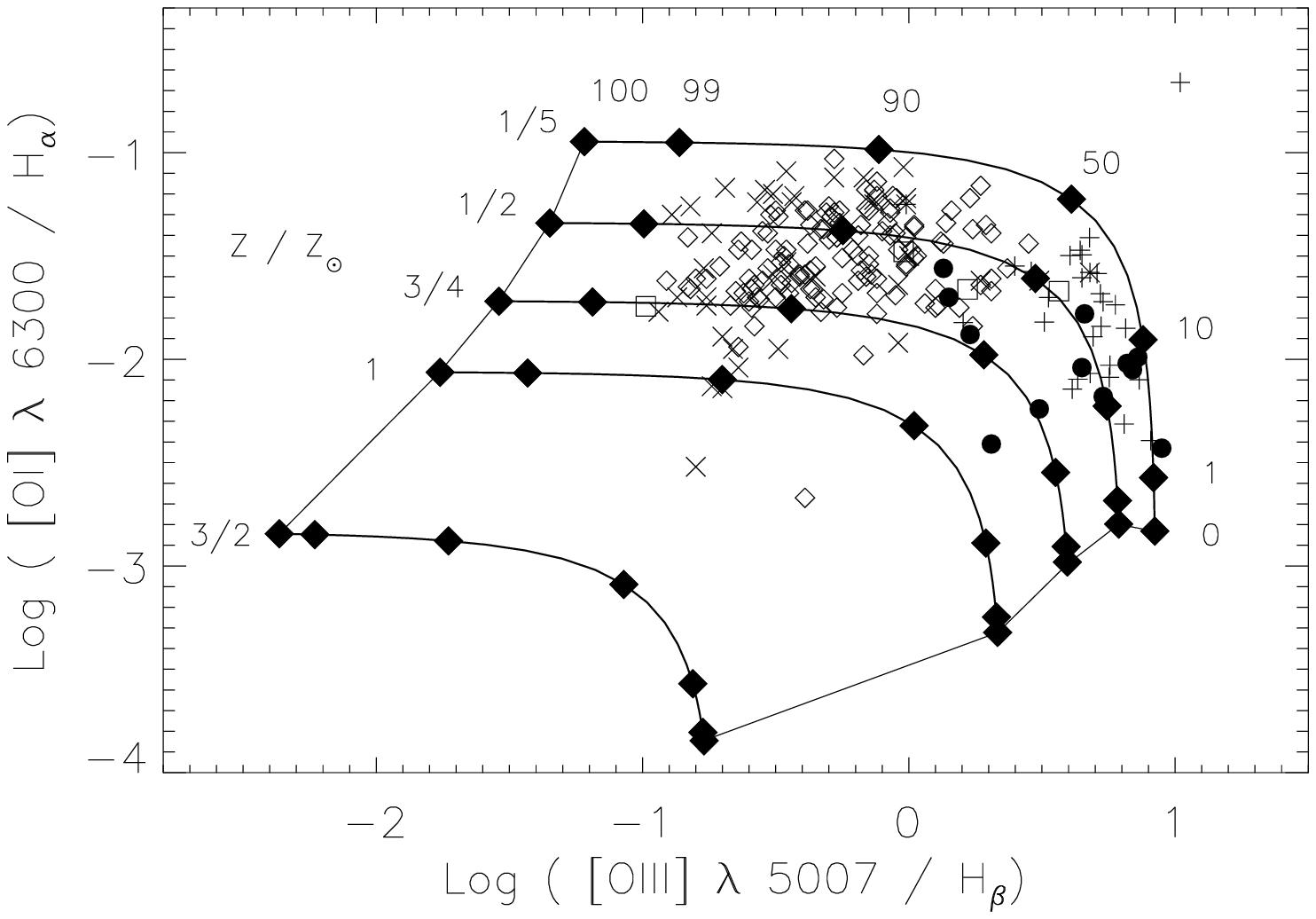,bbllx=40pt,bblly=10pt,bburx=500pt,bbury=312pt,width=8.5cm}
\psfig{figure=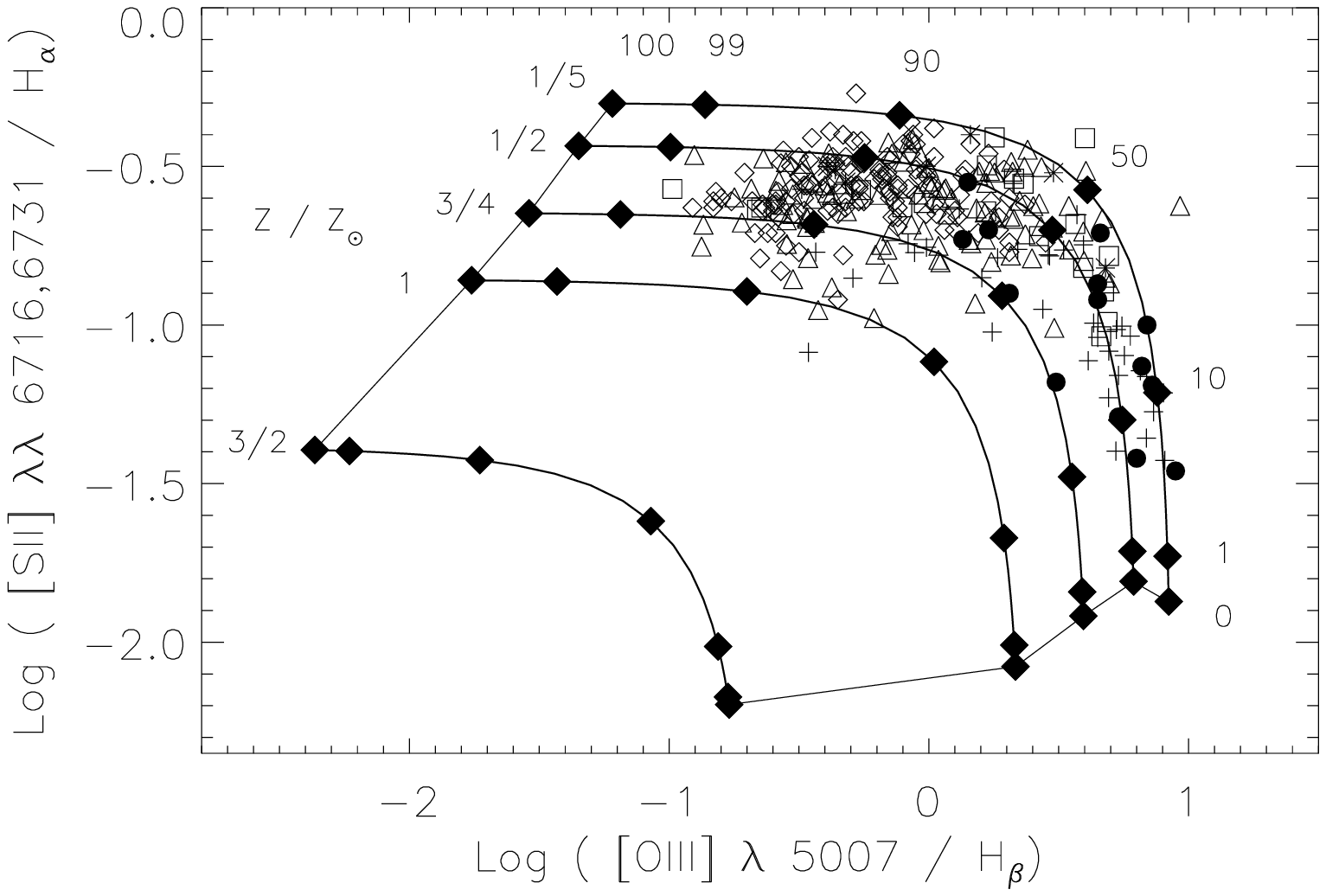,bbllx=40pt,bblly=10pt,bburx=500pt,bbury=312pt,width=8.5cm}}}
\parbox{15cm} {\caption{Comparison of the data with five sequences obtained by 
varying the relative weights of the high ($f=1$) and low ($f=10^{-5}$) 
excitation components. Solid diamonds are the points for which the low 
excitation component contributes to 0, 1, 10, 50, 90, 99 and 100\,\% of the 
H$\beta$ flux (from right to left). 
\label{Gridsrel} }} 
\end{center}
\end{figure*}

\begin{center}
\begin{table*}[hbtp]
\begin{center}
\begin{tabular}{|c|c|c|c|c|c|c|}
\hline
$n_\mathrm{H}$ & $f$ & $\log(U)$ & [\ion{O}{iii}]$_{\lambda 5007}$/H$\beta$ & 
[\ion{O}{i}]$_{\lambda 6300}$/H$\alpha$ & 
[\ion{S}{ii}]$_{\lambda\lambda 6716,6731}$/H$\alpha$ \\
\hline
10 cm$^{-3}$ & 1 & $-1.89$ & $7.53$ & $2.74$$\times 10^{-3}$ & 
2.39$\times 10^{-2}$ \\
& $10^{-1}$ & $-2.57$ & 5.88 & 8.47$\times 10^{-3}$ & 6.77$\times 10^{-2}$ \\
& $10^{-2}$ & $-3.24$ & 3.35 & 2.27$\times 10^{-2}$ & 1.69$\times 10^{-1}$ \\
& $10^{-3}$ & $-3.92$ & 1.04 & 5.13$\times 10^{-2}$ & 3.29$\times 10^{-1}$ \\
& $10^{-4}$ & $-4.6$1 & 0.17 & 9.17$\times 10^{-2}$ & 3.29$\times 10^{-1}$ \\
\hline
1000 cm$^{-3}$ & 1 & $-1.29$ & 9.29 & 8.88$\times 10^{-4}$ & 
7.11$\times 10^{-3}$ \\
& $10^{-1}$ & $-1.91$ & 8.27 & 2.90$\times 10^{-3}$ & 2.08$\times 10^{-2}$ 
\\
& $10^{-2}$ & $-2.57$ & 6.47 & 8.90$\times 10^{-3}$ & 5.93$\times 10^{-2}$ 
\\
& $10^{-3}$ & $-3.25$ & 3.66 & 2.39$\times 10^{-2}$ & 1.49$\times 10^{-1}$ 
\\
& $10^{-4}$ & -$3.92$ & 1.12 & 5.44$\times 10^{-2}$ & 2.93$\times 10^{-1}$ 
\\
\hline
\end{tabular}
\parbox[b]{11cm}{\caption{Variations of line ratios induced by a change of 
density. The age is 0\,Myr and the metallicity is $Z_{\odot}/5$.
\label{TAB_nH}}} 
\end{center}
\end{table*}
\end{center}
Models of pure instantaneous starbursts, with metallicities in the range
$[1/5,3/2]Z_{\odot}$, are compared to the data in Fig.~\ref{GridsfZ1} for six
values of the filling
factor $f$: $10^{-5}$, $10^{-4}$, $10^{-3}$, $10^{-2}$, $10^{-1}$, and 1. The slope of the IMF is -1.35, for a mass range of 0.1-120 M$_{\odot}$. The effects of dust are not taken into account in the models plotted on Fig.~\ref{GridsfZ1}.
The data are largely covered by the models when at least two parameters vary -- here
chemical and geometrical.
Interestingly, different types of objects seem to occupy 
different zones in the $Z$-$U$ plane: 
\ion{H}{ii} galaxies are in better agreement with low-$Z$ (1/50 $Z_{\odot}$ to
3/4 $Z_{\odot}$), high-$U$ models ($f\in[10^{-3},1]$); SBNGs are
compatible with low values of $f$ ($10^{-5}$ to $10^{-2}$), but encompass 
a wide range of metallicities. 

We also report on Fig.~\ref{GridsfZ2} the comparison of predictions with 
observations in following diagrams: [\ion{O}{ii}]$_{\lambda 3727}$/H$\beta$ vs. 
[\ion{O}{iii}]$_{\lambda 5007}$/H$\beta$ and [\ion{O}{ii}]$_{\lambda 3727}$/[\ion{O}{iii}]$_{\lambda 5007}$ 
vs. [\ion{S}{iii}]$_{\lambda 9069}$/[\ion{S}{ii}]$_{\lambda\lambda 6716,6731}$. Note that the 
[\ion{O}{ii}]/H$\beta$, [\ion{O}{ii}]/[\ion{O}{iii}] and [\ion{S}{iii}]/[\ion{S}{ii}] ratios {\em are} 
sensitive to reddening and, for this reason, we do not include them in the 
discussion. We can only notice that, given the uncertain
impact of dust on these ratios, our results are compatible with the 
observations. 
\subsection{The apparent relation between $Z$ and $U$: degeneracy problems}
\begin{center}
\begin{figure*}[thbp]
\begin{center}
\centerline{\hbox{
\psfig{figure=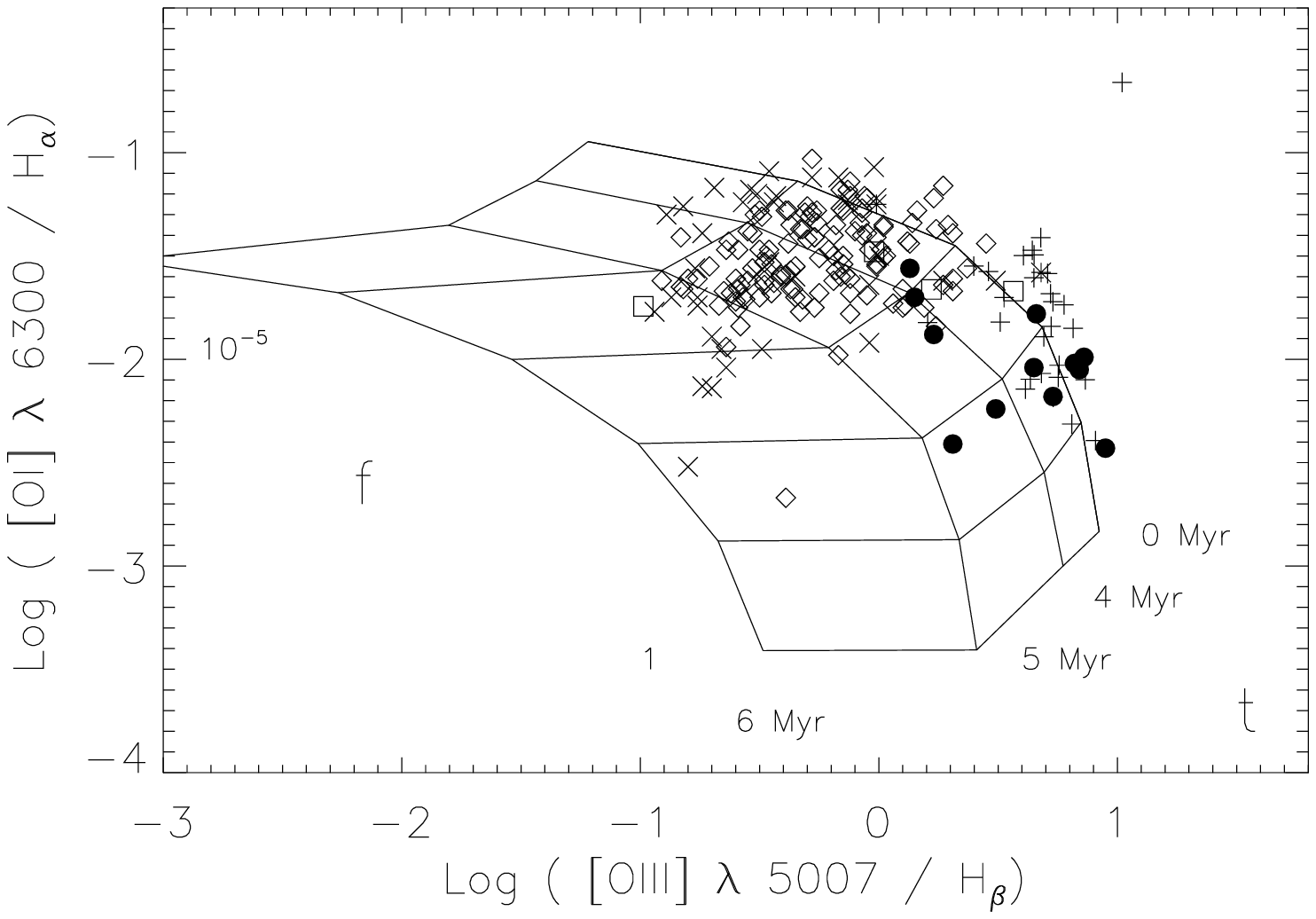,bbllx=40pt,bblly=10pt,bburx=480pt,bbury=312pt,width=7.5cm}
\psfig{figure=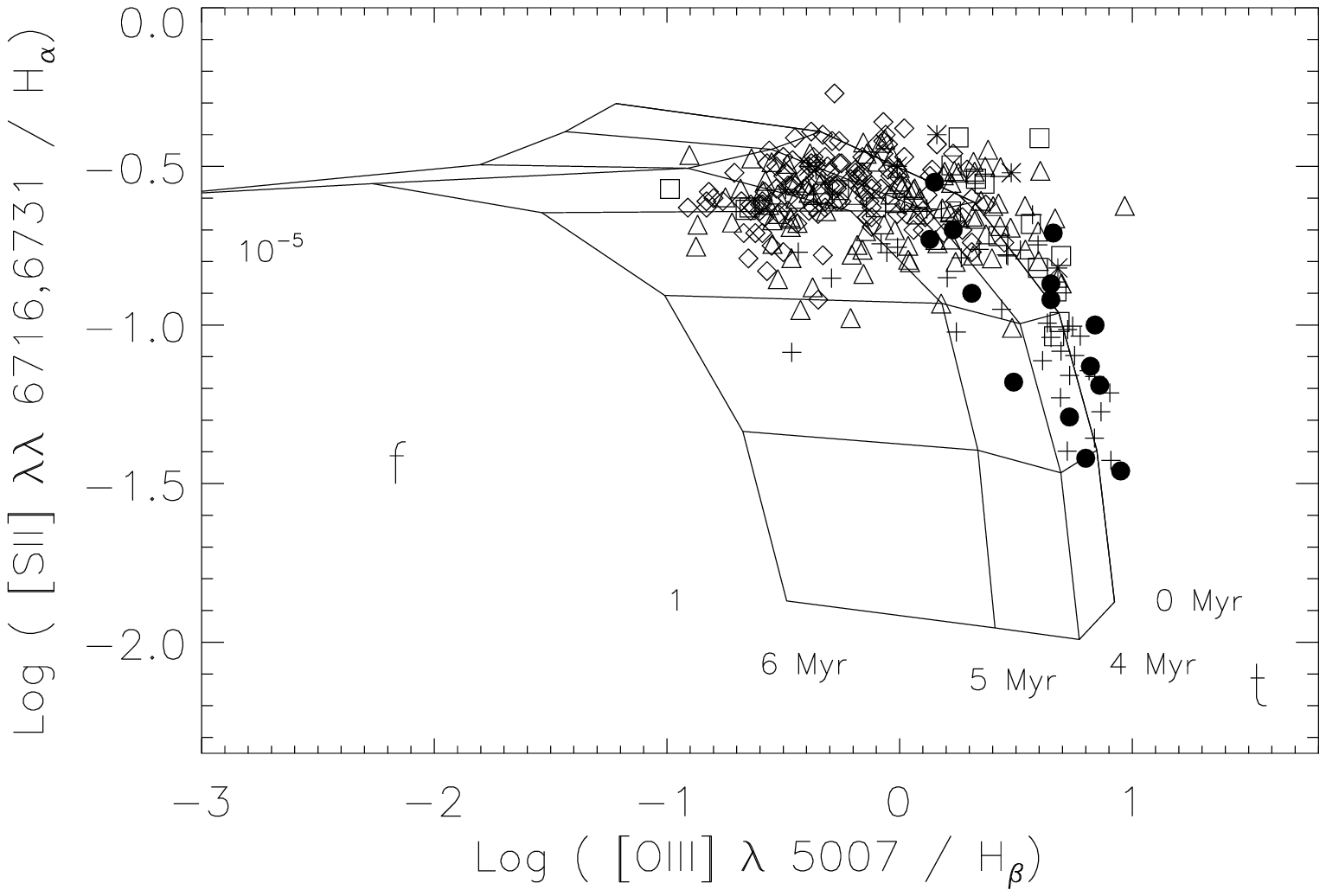,bbllx=10pt,bblly=10pt,bburx=450pt,bbury=312pt,width=7.5cm}}}
\parbox{11cm} {\caption{Evolution with the starburst age of the 
$Z_{\odot}/5$-model sequence plotted on Fig.~\ref{GridsfZ1} \label{DEG_t} }} 
\end{center}
\end{figure*}
\end{center}

\begin{figure}[thbp]
\begin{center}
\centerline{\hbox{
\psfig{figure=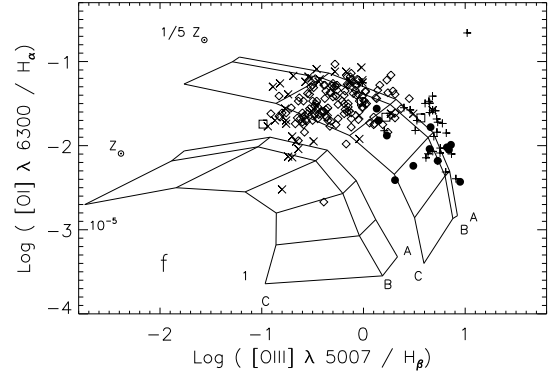,bbllx=40pt,bblly=10pt,bburx=490pt,bbury=335pt,width=7.5cm
}
}}
\parbox{8cm}{\caption{Impact of the IMF on [\ion{O}{i}]/H$\alpha$ vs. [\ion{O}{iii}]/H$\beta$ for two metallicities ($Z_{\odot}/5$ and $Z_{\odot}$). In each 
case, three $f$ sequences are plotted: A ($\alpha=-2.35$, 
$M_\mathrm{low}=0.1\,M_{\odot}$ and $M_\mathrm{up}=120\,M_{\odot}$), B ($\alpha=-3.35$, $M_\mathrm{low}=0.1\,M_{\odot}$ and 
$M_\mathrm{up}=120\,M_{\odot}$) and C ($\alpha=-2.35$, $M_\mathrm{low}=3\,M_{\odot}$ and $M_\mathrm{up}=30\,M_{\odot}$). \label{DEG_IMF}}} 
\end{center}
\end{figure}

A mean $U$-$Z$ relation fitting the data could easily be derived 
from Fig.~\ref{GridsfZ1}, as in Dopita \& Evans (1986). 
However, this relation would suffer from too many uncertainties: first, 
it would 
not take into account the high $Z$-low $Z$ degeneracy; second, age effects are 
very similar to those induced by the metallicity. This is illustrated in 
Fig.~\ref{DEG_t}, where [\ion{O}{i}]/H$\alpha$ and 
[\ion{S}{ii}]/H$\alpha$ vs. [\ion{O}{iii}]$_{\lambda 5007}$/H$\beta$ are 
plotted for 
$Z_{\odot}/5$ at $t=0$, 4, 5 and 6\,Myr. 
Variations of the IMF give rise to similar uncertainties, as 
shown on Fig.~\ref{DEG_IMF}: the lower number of massive stars for a steeper or
truncated 
IMF causes a strong decrease of both [\ion{O}{iii}]/H$\beta$ and [\ion{O}{i}]/H$\alpha$, just as if the metallicity was higher. 

Due to this age-metallicity-IMF degeneracy, a meaningful $Z$-$U$ analytical 
relation cannot be established on the basis of line ratio diagrams. 
Moreover, Fig.~\ref{GridsfZ1} clearly shows that the {\em dispersion} of $U$, 
rather than the mean value of this parameter, is linked to the metallicity:
while starbursts exist at any $U$ at low $Z$, only low-$U$ starbursts 
are observed at high metallicity. We consider a possible cause of 
this trend in the next paragraph.
\subsection{The upper envelope of the data sequences}

\begin{figure}[htb]
\begin{center}
\centerline{\hbox{
\psfig{figure=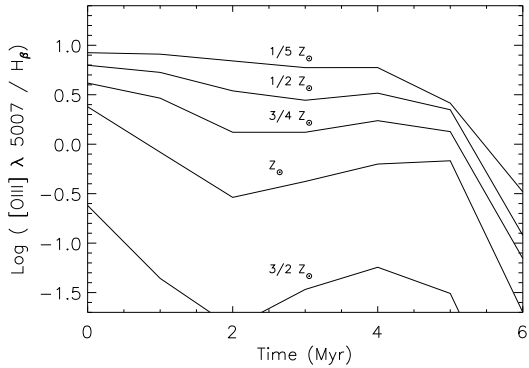,bbllx=40pt,bblly=10pt,bburx=470pt,bbury=360pt,width=6.5cm
}
}}
\parbox{8cm} {\caption{Time evolution of the [\ion{O}{iii}]$_{\lambda 5007}$/H$\beta$ ratio for instantaneous starbursts. The decrease of the ratio after 
5\,Myr is very rapid and puts severe constraints on the age of starbursts (see 
text for details) \label{Evtime} }} 
\end{center}
\end{figure}

In the [\ion{O}{i}]/H$\alpha$ vs. [\ion{O}{iii}]/H$\beta$ and [\ion{S}{ii}]/H$\alpha$ vs. [\ion{O}{iii}]/H$\beta$ diagrams, many data points lie above 
predictions, even when the misclassified objects of Leech et al. (1989) are 
excluded from the sample. To check whether this discrepancy can be 
solved by varying the density, we have performed two series of  
calculations at $Z=Z_{\odot}/5$ with $n_\mathrm{H}=10$ and 1000\,cm$^{-3}$. We chose 
this metallicity because the corresponding sequence defines the upper envelope of 
our model grid in these diagrams. The results are shown in Table~\ref{TAB_nH}. The main one is that model sequences with different densities 
largely overlap: clearly, the outliers
cannot be explained by a variation of $n_\mathrm{H}$. 

Many authors have 
proposed that a faint contribution from shocks or AGNs could explain  the 
intensity of the [\ion{O}{i}] line in starbursts (see e.g. Stasi\'nska \& Leitherer 1996). 
We consider here an alternative possibility where there are, inside the 
starburst, both high- and low-excitation zones, due for example to a distribution 
of gas and stars more complex than the classical geometry adopted in our 
calculations. To test this hypothesis, we computed a series of sequences 
obtained by varying the relative weight of high- and 
low-excitation models ($f=1$ and $f=10^{-5}$, respectively) for 
five metallicities. The results (Fig.~\ref{Gridsrel}) are in much better agreement with the data than pure $U$ 
sequences, as they explain both the trend and the dispersion of the 
observations. Another advantage of this explanation is that no other source of ionization, such as AGNs or shocks, is required. 
If this hypothesis is correct, the evolution of the apparent 
$U$ with $Z$ in fact corresponds to the presence in 
low-metallicity objects of a high-$U$ component,
which is absent in high-$Z$ starbursts. We come 
back to this point in the discussion.
\subsection{Constraints on age and IMF}

Fig.~\ref{Evtime} shows the evolution, up to 6\,Myr, of the [\ion{O}{iii}]$_{\lambda 5007}$/H$\beta$ ratio for the models plotted on Fig.~\ref{GridsfZ1} 
and \ref{GridsfZ2}. The constraints on age shown by this plot are 
very strong, though noticeably different for \ion{H}{ii}
galaxies and SBNGs: \ion{H}{ii} galaxies, of high-excitation level, need the high
energy photons emitted by massive stars, so that only young ($\leq$3\,Myr)
models are compatible with observations (see Sect.~7); SBNGs, requiring a
lower excitation level, span a wider range of ages. In any case, the 
rapid drop of the [\ion{O}{iii}]/H$\beta$
ratio after 5\,Myr requires the presence of stars younger than 6\,Myr.
\begin{figure*}[thbp]
\begin{center}
\centerline{\hbox{
\psfig{figure=
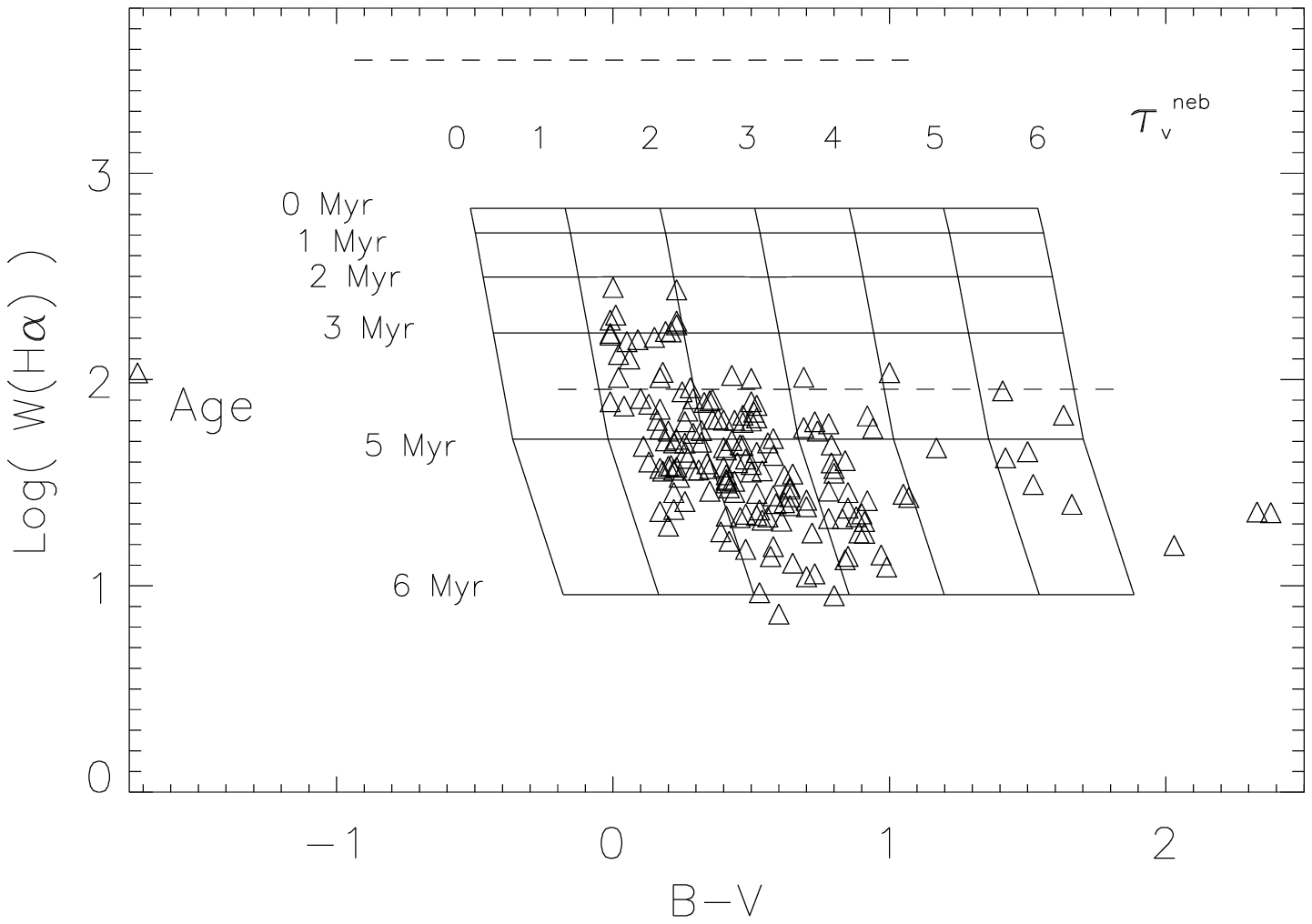,bbllx=20pt,bblly=10pt,bburx=480pt,bbury=312pt,width=8.5cm}
\psfig{figure=
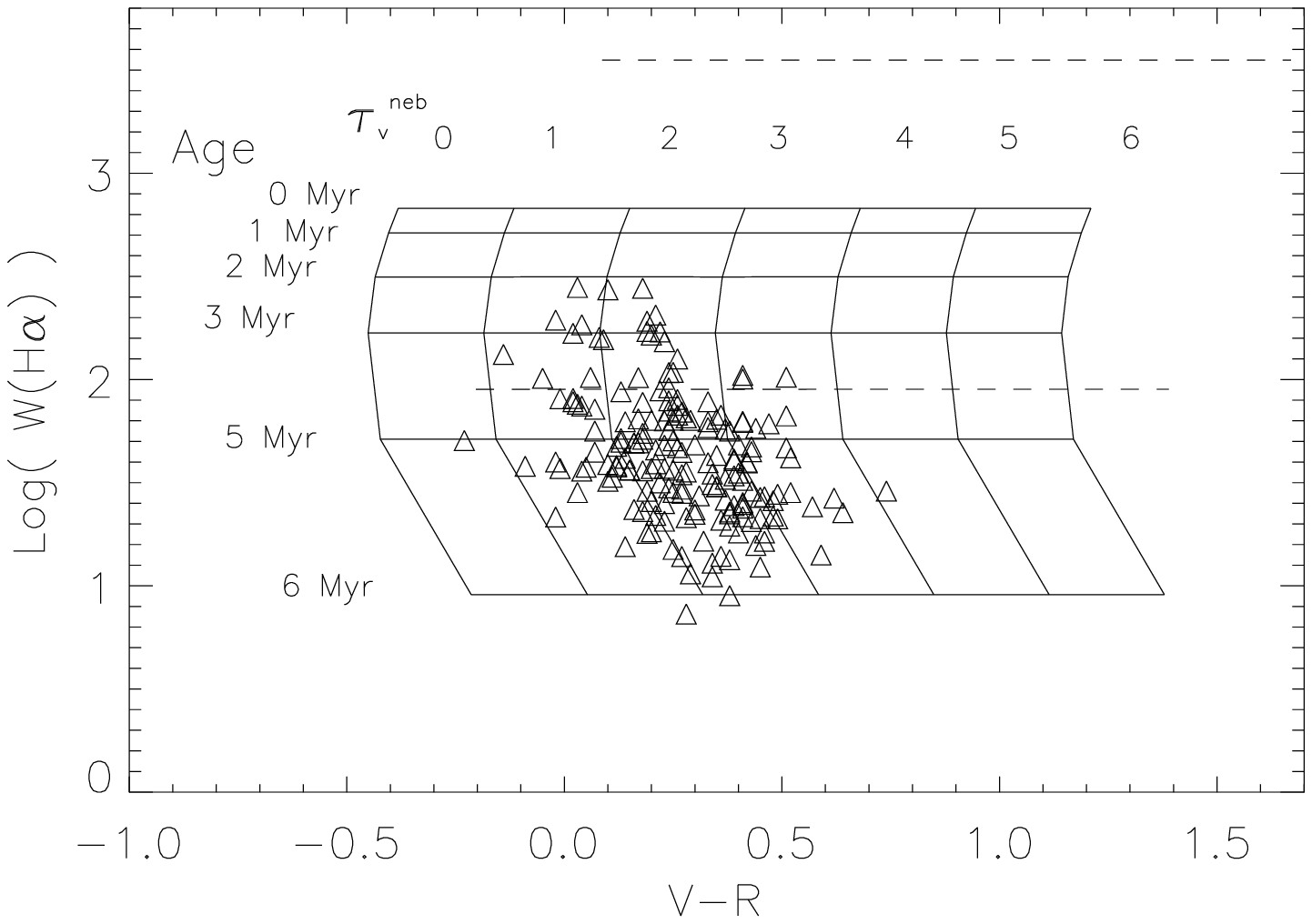,bbllx=20pt,bblly=10pt,bburx=480pt,bbury=312pt,width=8.5cm}
}} 
\parbox{17cm}{\caption{ Pure starburst models
for various ages (0 to 6\,Myr) and  $\tau_{V}$ between 0 to 6. The ratio $\tau_{V}^\mathrm{stell}/\tau_{V}^\mathrm{neb}=0.5$ is equal to 1. The covering 
factor is
0.1. Data (triangles) are from Contini et al. (1998). The solid lines 
indicate the predictions of the $\Omega/(4\pi)=0.1$ model grid. The two dashed 
lines show the $t=0$\,Myr (upper line) and $t=6$\,Myr (lower line) model results 
for $\Omega/(4\pi)=1$. 
\label{Wha_col1}}}
\end{center}
\end{figure*}
\begin{center}
\begin{figure*}[htbp]
\begin{center}
\centerline{\hbox{
\psfig{figure=
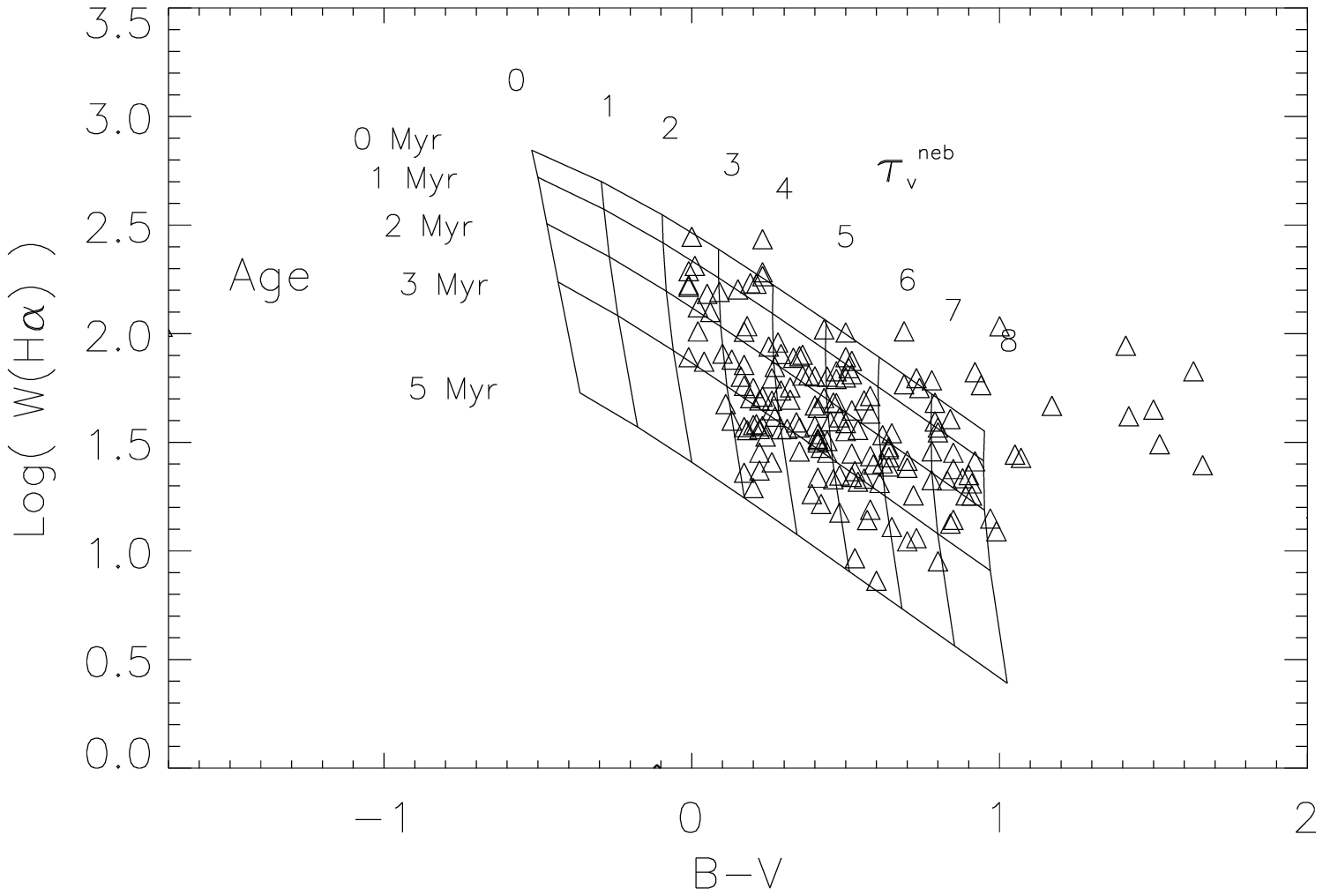,bbllx=20pt,bblly=10pt,bburx=480pt,bbury=312pt,width=8.5cm}
\psfig{figure=
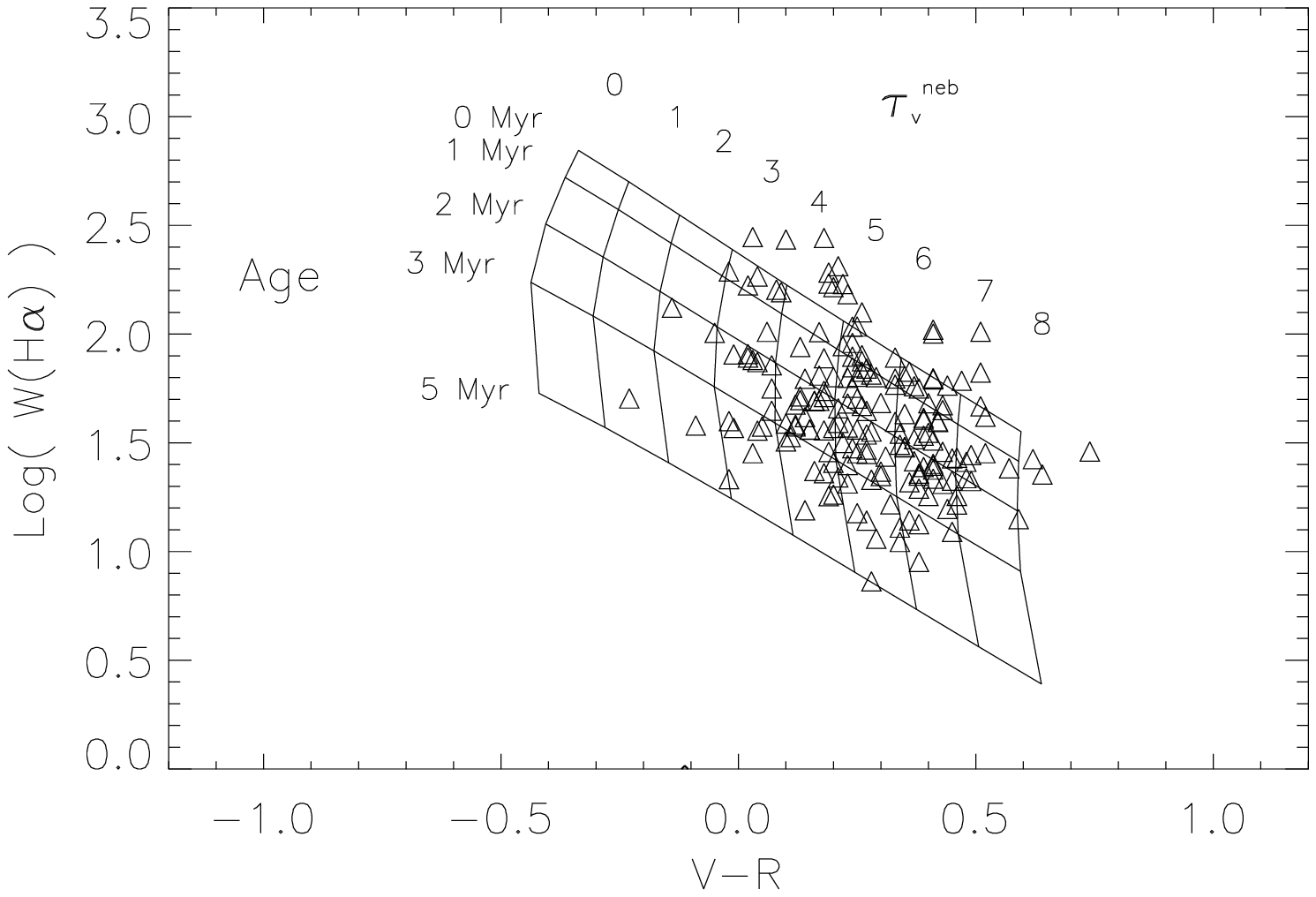,bbllx=20pt,bblly=10pt,bburx=480pt,bbury=312pt,width=8.5cm}
}}
\parbox[b]{9.8cm} {\caption{Same as Fig.~\ref{Wha_col1}, but for 
$\tau_{V}^\mathrm{stell}/\tau_{V}^\mathrm{neb}=0.5$ (see 
Sects. 3.3 and 5).
\label{Wha_col2}}}
\end{center}
\end{figure*}
\end{center}
Line ratios depend on the hardness of the spectrum, and so on the IMF.
The [\ion{O}{i}] line, emitted
from the partly ionized zone, requires high-energy photons and is thus 
particularly sensitive to the high-mass end of the IMF. Fig.~\ref{DEG_IMF} 
presents the sensitivity of [\ion{O}{i}]$_{\lambda 6300}$/H$\alpha$ and [\ion{O}{iii}]/H$\beta$ to the initial mass function. A Salpeter IMF is favored by our
results, in good agreement with recent studies (e.g. Stasi\'nska and
Leitherer 1996; Garc\'{\i}a-Vargas et al. 1995). An interesting point 
is that the steeper and truncated
IMFs, while clearly excluded by the data for high-$U$ \ion{H}{ii}Gs, 
are acceptable for metal-rich starbursts.

\section{Results on equivalent widths and colors: covering factor and 
extinction}

In the 
following, we aim to keep our previous fits of emission line ratios and, 
simultaneously, to reproduce $W(\mathrm{H}\alpha)$ and $B-V$ and $V-R$ colors from the 
sample of Contini et al. (1998). The models plotted on Fig.~\ref{GridsfZ1} and \ref{GridsfZ2} predict 
H$\alpha$ equivalent widths above 500\,\AA, while the observed
$W(\mathrm{H}\alpha)$ seldom exceed 300\,\AA.
Predicted equivalent widths in excess to the observations
are a well-known problem (Bresolin et al. 1999 and references therein). 
The hypothesis of a  ``density bounded'' nebula (Leitherer et al. 1996) is ruled out by the observed intensities of the low-ionization  lines; the reduction of the thickness of the partly-ionized zone would lead to
lower [\ion{S}{ii}]/H$\alpha$ and [\ion{O}{i}]/H$\alpha$  than 
in Fig.~\ref{GridsfZ1}, and the models would not fit any more these line ratios.

A low covering factor $\Omega/(4\pi)$ reduces the 
equivalent widths by increasing
the stellar emission relative to the nebular one, {\em without changing 
line ratios}. 
A similar explanation of  the 
faint $W(\mathrm{H}\beta)$ of blue compact and irregular 
galaxies was suggested by Mas-Hesse \& Kunth (1999). 

\begin{table*}[hbtp]
\begin{center}
\begin{tabular}{|c|c|c|c|ccccc|}
\hline $Z/Z_{\odot}$& $t_\mathrm{gal}$ & $\nu \times 10^{-3}$ 
&$t_\mathrm{c}$ & &&\%
(host)& (d)& \\ &(a) & (b) & (c) &A&B&C&D&E\\ \hline 1/5 & 11000 &
0.02 & 5000 & 100.0 & 21.0 & 11.7 & 2.5 & 0.0 \\ 1/2 & 11000 & 0.05 &
5000 & 100.0 & 17.9 & 9.6 & 2.5 & 0.0 \\ 3/4 & 13000 & 0.05 & 2000 &
100.0 & 16.2 & 8.9 & 1.9 & 0.0 \\ 1 & 13000 & 0.1 & 2000 & 100.0 &
14.5 & 7.8 & 1.9 & 0.0 \\ 3/2 & 13000 & 0.15 & 2000 & 100.0 & 12.4 &
6.6 & 1.7 & 0.0\\ \hline 
\end{tabular}
\vskip 0.5cm
\parbox{14cm}{\caption{Parameter values of
starburst + host scenarios. (a): Host galaxy age (in Myr) when
starburst occurs. (b): star formation parameter $\nu$ (in Myr$^{-1}$) of 
the host galaxy. (c) Time-scale of
gas infall onto the host galaxy (Myr). (d): fraction of 
ionizing photons emitted by the host population when the starburst 
starts in models A, B, C, D and E (see text).}} 
\end{center}
\end{table*}

Fig.~\ref{Wha_col1} shows the influence of
dust and age in the $W(\mathrm{H}\alpha)$ vs. $B-V$ and $W(\mathrm{H}\alpha)$ vs. $V-R$ 
planes, for $\Omega/(4\pi)=0.1$. For comparison, we also show
$t=0$ and $t=6$\,Myr sequences 
for $\Omega/(4\pi)=1$. 
We plot only $Z=Z_{\odot}$ models, since the metallicity has 
only weak effects on $B-V$, $V-R$ and $W(\mathrm{H}\alpha)$. 
Similarly, for such a low value of $\Omega/(4\pi)$, $f$ hardly affects the 
colors 
and the only models on Fig.~\ref{Wha_col1} are for $f=1$. 
Ages between 0 and 6\,Myr, 
$\tau_{V}^{\mathrm{stell}}/\tau_{V}^{\mathrm{gas}}=1$ and $\tau_{V}^{\mathrm{gas}}$ in the range [0,6] are 
suggested by the $W(\mathrm{H}\alpha)$ vs. $B-V$ diagram. 
However, $\tau_{V} \geq 4$ is excluded by the $V-R$ data, and 
ages
$\geq 6$\,Myr are incompatible with line ratios (see also Sect.~7.2). 

A ratio $\tau_{\lambda}^{\mathrm{stell}}/\tau_{\lambda}^{\mathrm{gas}}\simeq 
0.5$ (following Calzetti et al. 1997; see also Fanelli
et al. 1988; Keel 1993; Mas-Hesse \& Kunth 1999) can simultaneously fit
$B-V$ and $V-R$, but requires very high extinctions, up to 
$\tau_{V}^{\mathrm{neb}}=8$ (Fig.~\ref{Wha_col2}), while Contini et al. 
(1998) derived 
a maximum value of $\sim$ 4 from the Balmer decrement. The $B-V$ of
the reddest objects are considered as anomalous 
(Contini, private communication) and are not reproduced by 
any model. 
\section{The contribution of an evolved population}

The distribution of the data in the $W(\mathrm{H}\alpha)$ vs. $B-V$ or $V-R$ diagrams is very 
difficult to explain with pure instantaneous starburst models. The main problem 
is the very low covering factor required. A value of 0.1 seems in contradiction with previous estimates of the escaping fraction of ionizing ohotons in HII regions (see Sect. 7.2).
An alternative hypothesis, able to explain the 
equivalent widths and colors of the Contini et al. (1998) sample
with the much more reasonable value 
$\Omega/(4\pi)=0.5$, is the presence 
of evolved stars inside the starburst. 
Such a population, created during past formation episodes, can account for 
the observed weakness of equivalent
widths by increasing the continuum (e.g. McCall et 
al. 1985; D\'{\i}az et al.
1991). It may also explain the red colors of the sample without requiring huge 
amounts of dust. 

\begin{center}
\begin{figure*}[thbp]
\begin{center}
\centerline{\hbox{
\psfig{figure=
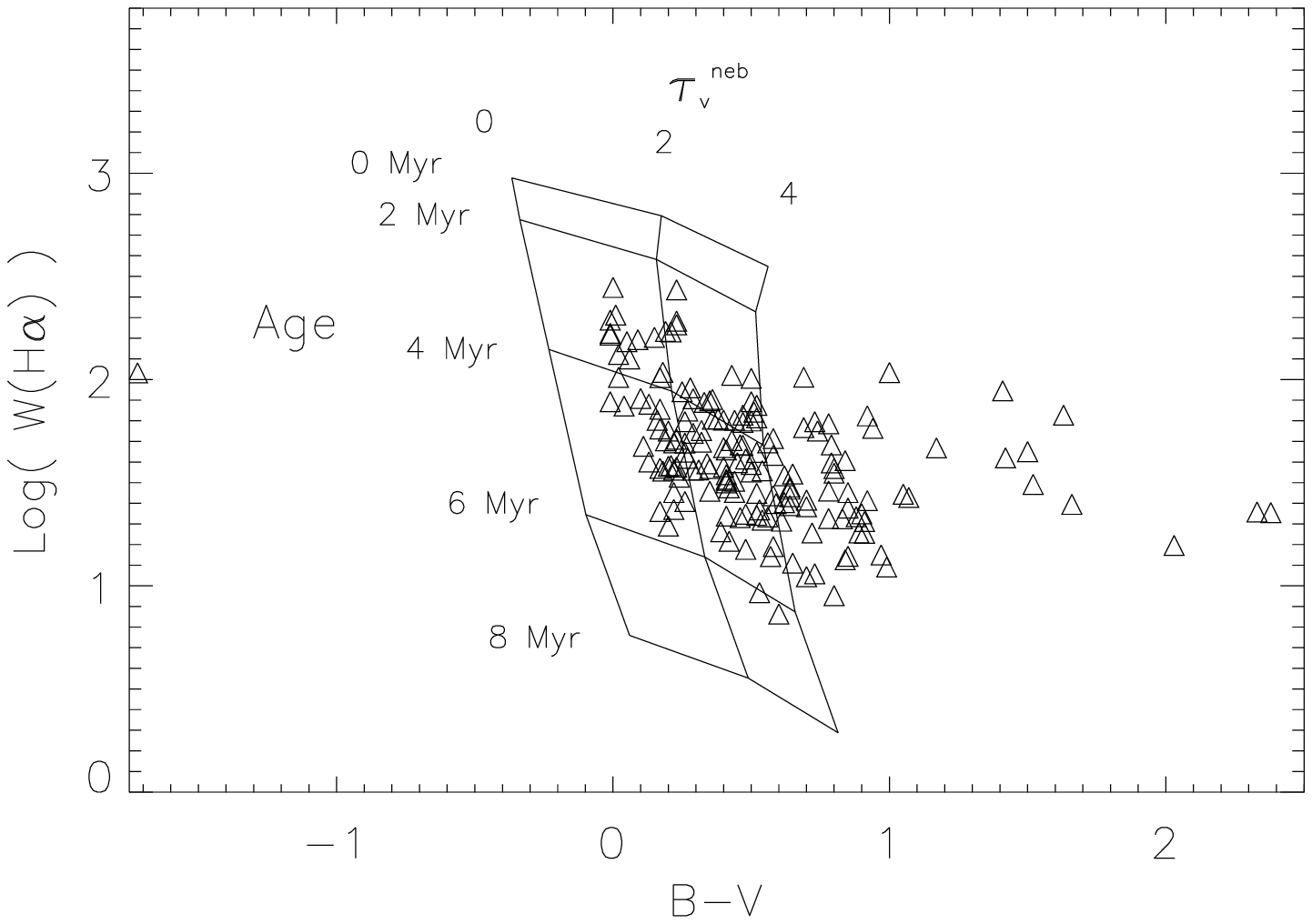,bbllx=20pt,bblly=10pt,bburx=480pt,bbury=312pt,width=8.5cm}
\psfig{figure=
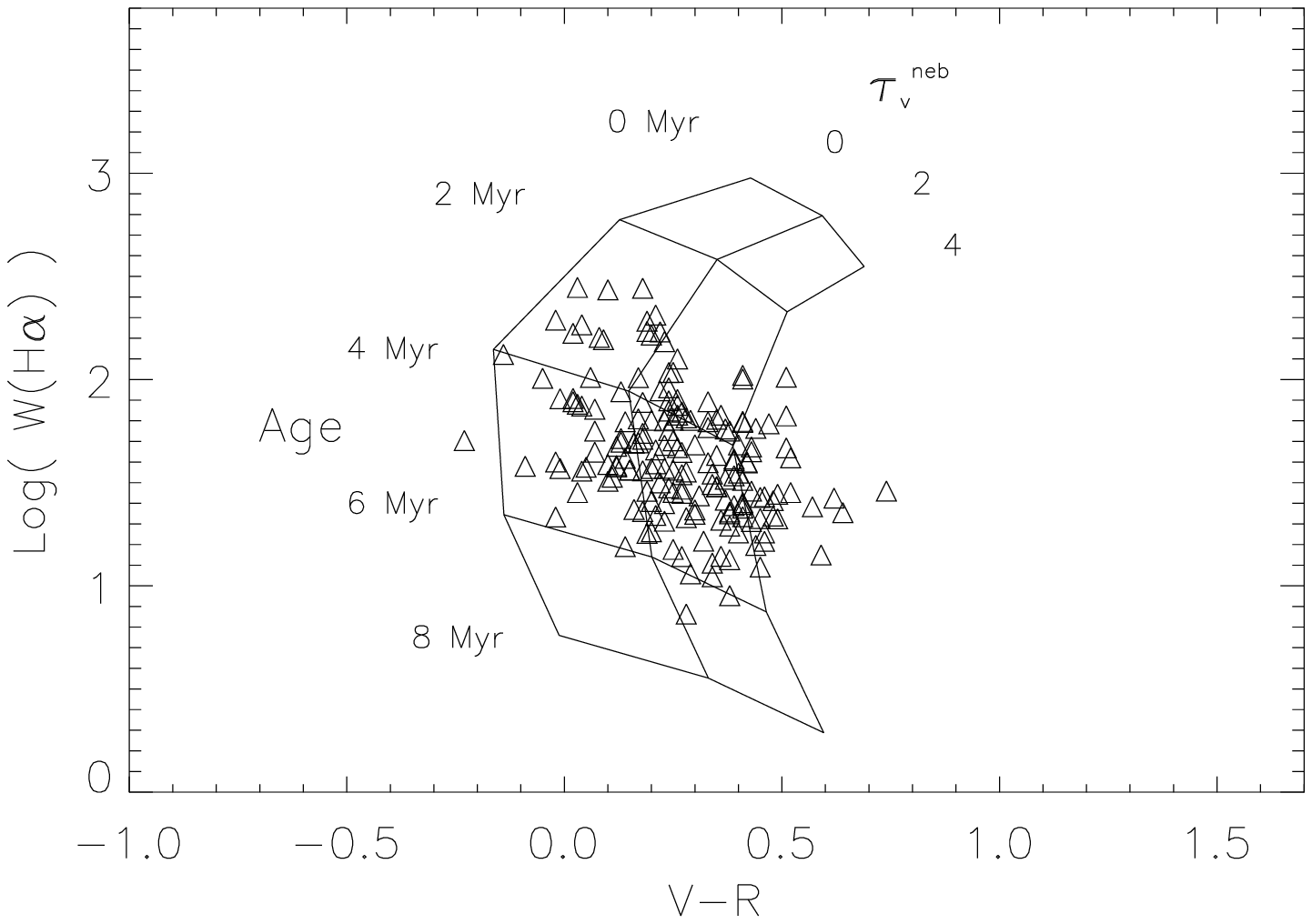,bbllx=20pt,bblly=10pt,bburx=480pt,bbury=312pt,width=8.5cm}
}} 
\parbox[b]{14cm} {\caption{H$\alpha$ equivalent widths and colors for 
instantaneous + 100\,Myr-old starbursts. The metallicity of the young starburst 
is solar. The old burst is assumed to have been ten times more intense 
than the current one. The model sequences are labeled in 
terms of the age of the present starburst and of the value of 
$\tau_{V}^\mathrm{neb}$. \label{Wha_col3}}}
\end{center} 
\end{figure*}
\end{center}

We considered two possible underlying populations. The first one consists of
an instantaneous burst which occured 100\,Myr ago. This old burst 
is assumed to have been ten times more intense than the current one. 
The predictions of these 
models are plotted on Fig.~\ref{Wha_col3}. As in Fig.~\ref{Wha_col2}, 
$\tau_{\lambda}^\mathrm{stell}/\tau_{\lambda}^\mathrm{neb}$ is set to 
0.5. However, these models face two problems: first, it is still necessary 
to extend the age of the young burst to $t=6$\,Myr, unless the covering factor 
is much lower than the value adopted on Fig.~\ref{Wha_col3} (0.5); second, the 
sum of the evolved and the young population does not reproduce the colors. To 
reconcile this type of models with the data, it would be necessary to assume 
that the old burst was systematically $\sim 100$ times more intense than the 
current one. 

The second type of underlying population is assumed to have formed  continuously. We computed a series of scenarios matching the observations of standard
Hubble sequence galaxies at $z=0$, and added a starburst. For the sake of 
consistency, the 
initial metallicity of the burst is 
the current metallicity $Z(t)$ of the host galaxy and is provided by P\'EGASE. 

The input parameters are chosen so that the starburst metallicities belong to 
the range $[1/5 Z_{\odot}, 3/2 Z_{\odot}]$ derived from line ratios. Each scenario is defined by three parameters:
the age $t_\mathrm{gal}$ of the host galaxy  when the starburst occurs; the 
infall timescale $t_\mathrm{c}$ used to parameterize 
the accretion rate of the galaxy; the gas-to-star conversion 
efficiency, $\nu$, relating the star formation rate $\zeta(t)$ to the gas 
density $\sigma(t)$ by the Schmidt law $\zeta(t)=\nu \sigma(t)$.

Five models (A, B, C, D, E) were computed for each scenario, 
corresponding to five different contributions of the evolved 
population to the total emission: A is the pure host galaxy, 
E is the pure starburst 
and B, C  and D are intermediate cases. Note that the 
star formation does not stop after the burst, but returns to the quiescent 
regime. The fraction of ionizing photons due to the underlying population 
immediately after the burst is given in Table 5, as well as the values of 
$t_\mathrm{gal}$, $t_\mathrm{c}$, and $\nu$ for the considered scenario.

\begin{center}
\begin{figure*}[thbp]
\begin{center}
\centerline{\hbox{
\psfig{figure=
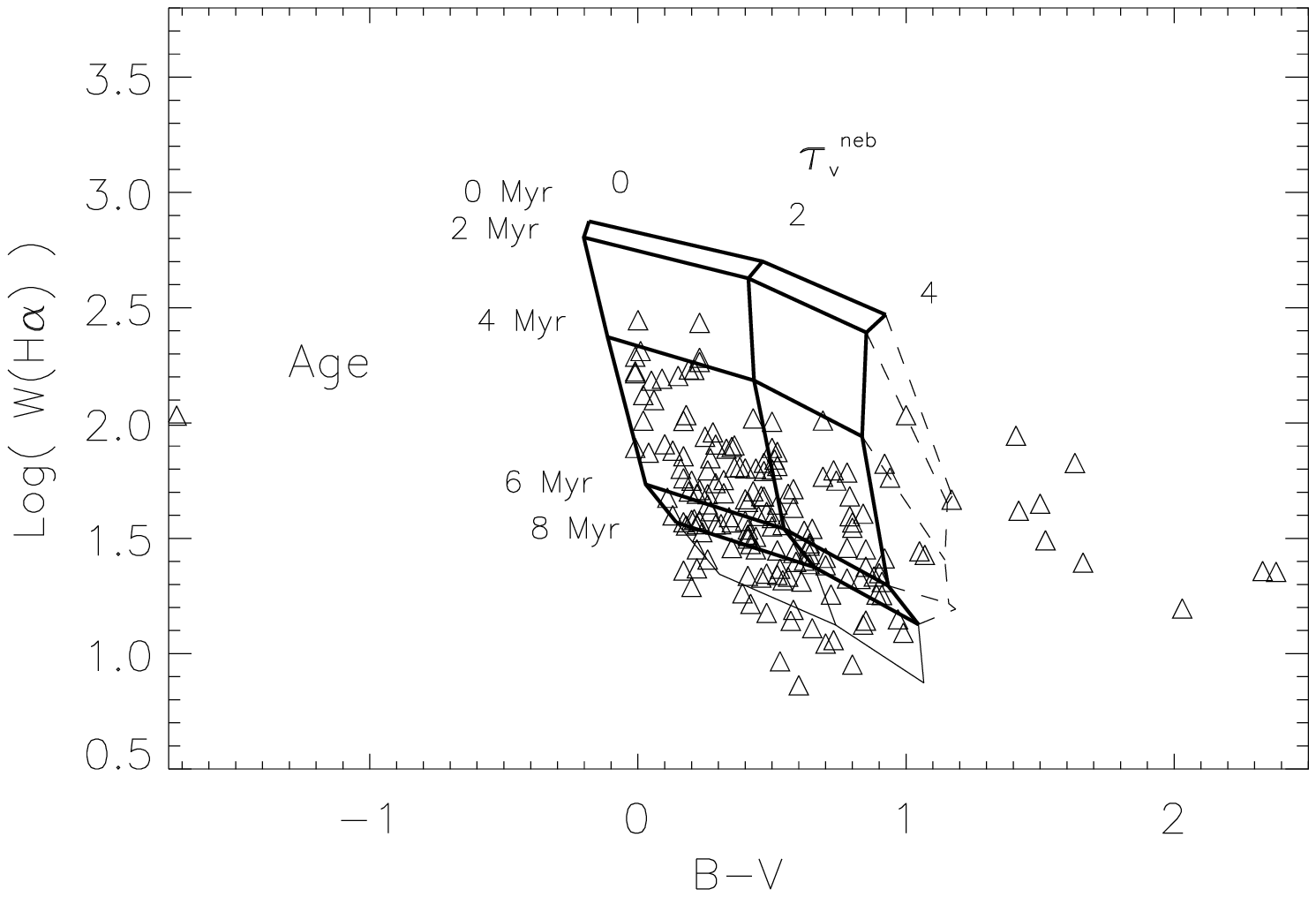,bbllx=20pt,bblly=10pt,bburx=480pt,bbury=312pt,width=8.5cm}
\psfig{figure=
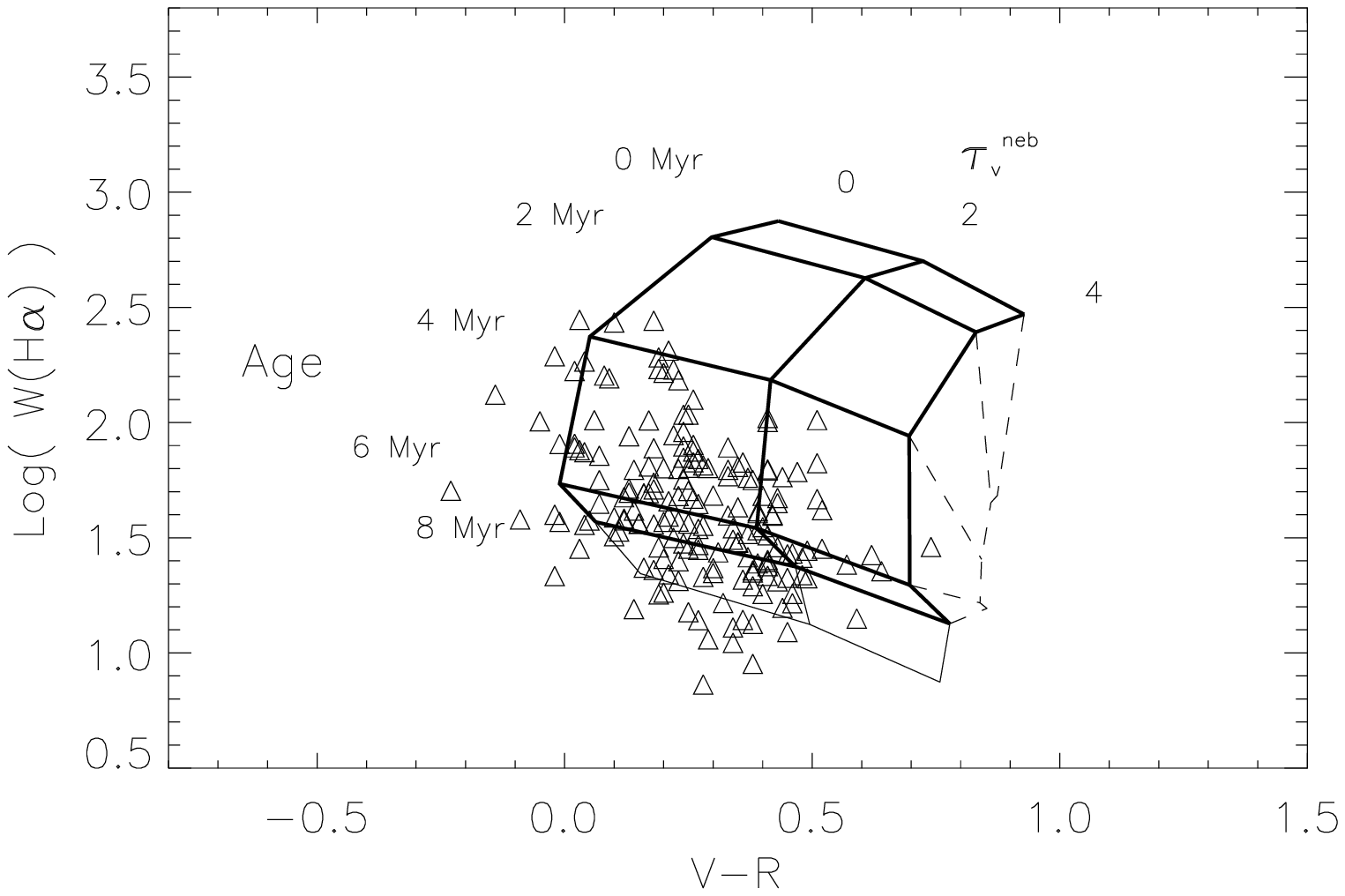,bbllx=20pt,bblly=10pt,bburx=480pt,bbury=312pt,width=8.5cm}
}} 
\parbox[b]{14cm} {\caption{H$\alpha$ equivalent widths and colors for 
instantaneous burst + continuous star formation scenarios. The thick solid 
lines show the predictions of model D for different ages and extinction levels 
of the burst. The metallicity of the burst is 1/2 $Z_{\odot}$. The dashed lines 
indicate the shift of the right part of the grid when the contribution of the 
underlying population is increased (model C). The thin solid lines show the 
shift of the low part of the grid when the metallicity is equal to 3/2 
$Z_{\odot}$.
\label{Wha_col4}}}
\end{center} 
\end{figure*}
\end{center}

Fig.~\ref{Wha_col4} presents the impact of the parameters on colors and equivalent widths. We compare the results of 
model D at $Z=Z_{\odot}/2$ to the sample
of Contini et al. (1998). We also show the results of model C for 
$Z=1/2 Z_{\odot}$, and the predictions of model D for 
 $3/2 Z_{\odot}$.
The filling factors are 
$10^{-2}$ at $1/2 Z_{\odot}$ and $5\times 10^{-4}$
at $3/2 Z_{\odot}$. As in Fig.~\ref{Wha_col2}, $\tau_{\lambda}^{\mathrm{stell}}$/$\tau_{\lambda}^{\mathrm{neb}}$ is set to 0.5. However, the covering factor (0.5) is 
significantly higher, and the $\tau_{V}^{\mathrm{neb}}$ range (0 to 4) is much more 
reasonable. 
Another important point is that, contrary to the pure starburst case, ages of 6\,Myr and more are compatible with emission line ratios. This is due to the 
presence of massive stars, formed after the burst in the underlying population. The presence of such stars ensures that the line ratios are virtually unchanged compared to their values during the 
burst. Contrary to the case of a burst that occured 100 Myr ago, these scenarios are able to explain simultaneously  the colors, the equivalent widths and the line ratios in our sample. They do not require very high levels of extinction, and, maybe more important, do not imply a covering factor lower than 0.5. For all these reasons,  the hypothesis of an underlying population formed continuously is our preferred one.  
\section{Discussion}
\subsection{A physical link between the ionization parameter and the metallicity?}
Dopita \& Evans (1986) proposed an analytical relation between the metallicity 
and the ionization parameter to describe the observational 
distribution of line ratios in their sample of extragalactic \ion{H}{ii} regions. In 
Sect.~4.3, we saw that such a relation suffers from many degeneracies due 
to age, IMF and metallicity. Moreover, it seems that the quantities which are 
really related are the dispersion of $U$ and the metallicity.

What could be the physical link between $Z$ and $U$? According to one 
hypothesis, this is a link through geometrical effects. 
Following Castor et al. (1975) and assuming their equations are valid also for a 
star cluster, the inner radius $R$ of 
a gaseous shell surrounding a massive star or cluster is: 
\begin{equation}
R \propto \left(\frac{\dot E t^{3}}{\rho} \right)^{1/5}, 
\end{equation} 
where $\dot{E}$ is the energy deposition rate of a star (or cluster) of age $t$ 
in the interstellar medium (ISM) of
density $\rho$. Leitherer et al. (1992) found that
$\dot{E}$ is constant for an instantaneous burst  until $\sim$ 6\,Myr 
and scales linearly with $Z$,
regardless of the age. Hence, high-$Z$ \ion{H}{ii} regions expand faster.

We therefore propose the following scenario to explain the distribution 
of starbursts in the line ratio diagrams. For some reason, a site of intense 
star formation appears in a galaxy with a metal-poor interstellar medium. This 
starburst can be composed of many individual \ion{H}{ii} regions surrounding star 
clusters. As these sub-components evolve with age, the \ion{H}{ii} regions expand 
because of stellar winds. At the same time, the enrichment 
of the ISM by massive-star ejecta begins. If star formation goes on, 
very young components will coexist inside the starburst with more evolved 
components. The former will dominate the nebular emission during the first 
few Myr, 
and the latter after.

Two additional effects can explain the lack of high-$Z$ and high-$U$ objects: 
first, at high metallicity, individual \ion{H}{ii} regions expand faster, leading 
to a lower ionization parameter; second, according to this scenario, the bulk 
of nebular emission in high-$Z$ starbursts is produced by evolved, 
low-excitation \ion{H}{ii} regions. Checking the validity of this hypothesis 
requires to model also the expansion of nebulae. This is 
the subject of a forthcoming paper. 
\subsection{The equivalent-width problem}
The problems encountered by pure starburst scenarios to account for the 
weakness of the H$\alpha$ equivalent widths from the sample of Contini et al. 
(1998) are several. The main one is that a very low covering factor is required. From Fig.~\ref{Wha_col1} and \ref{Wha_col2}, we deduce that a value 
$\Omega/(4\pi) \sim 0.1$ is necessary to reproduce both $W(\mathrm{H}\alpha)$
and the line ratios in the case of a pure starburst. This means that 90\,\% of the Lyman continuum 
photons do not contribute to ionization, but are directly absorbed by dust or 
escape from the starburst. This is in contradiction with many previous studies 
using H$\alpha$ or H$\beta$ luminosities to estimate the mass of stellar 
clusters inside starbursts. In many cases, the agreement between such 
estimations and those based on ultraviolet continuum luminosities is excellent 
(Gonz\'alez-Delgado et al. 1999). Some discrepancies appear frequently 
(Leitherer et al. 1996; Vacca et al. 1995), but they are however weak enough
to exclude an absorption fraction of the ionizing photons by the gas as low as 
10\,\%. Moreover, observations of diffuse H$\alpha$ emission in nearby 
galaxies lead to an estimation of 15--50\,\% for the fraction 
of ionizing photons escaping from individual \ion{H}{ii} regions 
(Kennicutt 1998 and references therein). 

Such a systematic low value of the covering factor is therefore clearly ruled 
out by the observations. Note also that pure starbursts models, even with 
$\Omega/(4\pi)=0.1$, cannot reproduce simultaneously the emission line 
ratios and the equivalent widths (see 
Sect.~5). Hence, the presence of an evolved stellar
population coexisting with a starburst, as analysed in Sect. 6, seems mandatory to explain
equivalent widths and colors, even for the data of Contini et al. (1998),
who strictly limitated their apertures to H$\alpha$-emitting 
areas. A similar conclusion for galaxies
observed through 5-arcsec. apertures was drawn
by Lan\c{c}on \& Rocca-Volmerange (1996)
from the near-infrared spectral synthesis of starbursts.
\subsection{Nature of the underlying population}
Including underlying populations in the models allows to fit the 
data in a satisfactory way in the $W(\mathrm{H}\alpha)$ vs. $B-V$ 
and $V-R$ diagrams with 
$\Omega/(4\pi)=0.5$, which means that $\sim 50\,\%$ of the ionizing photons escape 
from the gaseous nebula. This value is close to the 60\,\% fraction suggested by 
Mas-Hesse \& Kunth (1999). The nature of this population remains however 
elusive. 

We have tested an old (100\,Myr old) instantaneous burst as a possible 
underlying population. The agreement with the data, at first sight, is 
relatively good (Fig.~\ref{Wha_col3}). In this case, however, the age of the 
youngest burst has to extend up to 8\,Myr to explain the faint end of the 
W(H$\alpha$) observational distribution. At this age, our models are 
incompatible with line ratio data. Note that there is no Wolf-Rayet(W-R) 
star SED in our spectral library; instead, the spectra of the 
hottest stars available in the library are used during W-R stages.
The implementation of W-R SEDs in the library could in principle extend 
the range of ages compatible with line ratios up to 6\,Myr or more. As an 
example, Gonz\'alez-Delgado et al. (1998) obtained an age between 6 and 9\,Myr for 
IRAS 0833+6517. However, this conclusion is based on models with 
an IMF truncated at $M_\mathrm{up}=30\,M_{\odot}$. Such a low value is 
clearly excluded for the bulk of our sample (Fig.~\ref{DEG_IMF}). Moreover, the 
role of W-R stars in
starburst nebular emission is still under
debate. According to the most recent studies, the bulk of nebular emission 
is due to stellar populations younger than 3\,Myr (Bresolin et al. 1999). 

We have assumed that the old burst was only ten times more intense 
than the current one. Changing this value to one hundred, for example, would 
move the predictions on Fig.~\ref{Wha_col3} toward the bottom-right 
and might explain 
the bulk of the data with ages lower than, or equal to, 4\,Myr. In this case, 
however, such a difference between past and present star formation remains 
to be explained. 

In this context, the models plotted on Fig.~\ref{Wha_col4} appear as the most 
attractive ones. Scenarios with an underlying population providing from
2.5 up to 20\,\% of the number of
ionizing photons are in acceptable agreement with the data and 
explain naturally all the observables 
(emission lines, colors
and equivalent widths). The range of $\tau_{V}$  is consistent with the Balmer 
decrement measurements of Contini et al. (1998). Moreover, the age
and metallicity of the underlying population are typically those
of normal spirals, suggesting that starbursts are normal events in galaxy 
evolution. These scenarios draw a coherent picture of the starburst phenomenon, 
and for this reason, remain the most plausible ones. 
\section{Conclusion}
We have analyzed the emission line and continuum properties of a large sample of 
starbursts and put new constraints on their stellar populations,
metallicity, photoionization state and dust content. Our tool was the 
spectral evolution code P\'EGASE coupled to the photoionization 
code CLOUDY. 

From the comparison of a dataset of emission lines with pure instantaneous 
starburst scenarios, we have been able to constrain the evolution of the 
ionization parameter $U$ with the metallicity $Z$. At high abundance, the 
high-excitation component clearly detected at low $Z$ is absent or very weak. A 
preliminary analysis based on energy deposition rate considerations is 
unsufficient to explain this trend, and a more refined analysis 
of the dependency of the \ion{H}{ii} region expansion with $Z$ is needed.

The second important result concerns underlying populations 
inside starbursts; they are detected even through small-aperture 
observations. This contribution is needed to reproduce simultaneously all 
the observables of our sample
(line ratios, colors and equivalent widths). The underlying populations are 
simulated through the whole computation of the past star formation history of 
the galaxy hosting the starburst. In the scenarios favored by our analysis, 
the 
burst itself is supposed to be a brief, intense episode, preceded and followed 
by a quiescent regime. After the burst, this quiescent formation rate is 
sufficient to keep the line ratios compatible with the data. 

Future prospects should consider the problem of metal depletion on dust grains,
since depletion changes the composition of nebular gas. However, the question 
of the presence of dust \emph{inside}
\ion{H}{ii} regions will be solved only with more constraining data, 
e.g. those of the ISO satellite. 
Finally, the large grids computed with the help of
a coupled photoionization and stellar evolutionary synthesis
model will be also used to study other samples with better 
statistics.
\begin{acknowledgements}
We are pleased to thank Gary Ferland for his multiple consultation 
and help for the interface with CLOUDY. M.~F. acknowledges support from the 
National Research Council through the Resident Research Associateship Program. 
\end{acknowledgements}
\bibliography{} 

\end{document}